\documentclass{emulateapj}

\newcommand{\citepeg}[1]{\citep[{e.g.,}][]{#1}}

\shorttitle{Radio Emission from GRB Host Galaxies}
\shortauthors{Perley et al.}

\begin{document}

\title{Connecting GRB\lowercase{s} and ULIRG\lowercase{s}:  A Sensitive, Unbiased Survey for Radio Emission from Gamma-Ray Burst Host Galaxies at \lowercase{$0<z<2.5$}}

\def\msol{${\rm M}_\odot$}

\def\cit{1}
\def\hubble{2}
\def\nrao{3}
\def\dark{4}
\def\edinburgh{5}
\def\gsfc{6}
\def\jssi{7}
\def\iceland{8}
\def\eso{9}
\def\warwick{10}
\def\leicester{11}
\def\mail{*}

\author{D.~A.~Perley\altaffilmark{\cit,\hubble,\mail},
        R.~A.~Perley\altaffilmark{\nrao},
        J.~Hjorth\altaffilmark{\dark},
        M.~J.~Micha{\l}owski\altaffilmark{\edinburgh},
        S.~B.~Cenko\altaffilmark{\gsfc,\jssi}, \\
        P.~Jakobsson\altaffilmark{\iceland},
        T.~Kr\"uhler\altaffilmark{\eso},
        A.~J.~Levan\altaffilmark{\warwick},
        D.~Malesani\altaffilmark{\dark}, and
        N.~R.~Tanvir\altaffilmark{\leicester}
}

\altaffiltext{\cit}{Department of Astronomy, California Institute of Technology,
MC 249-17,
1200 East California Blvd,
Pasadena CA 91125, USA}
\altaffiltext{\hubble}{Hubble Fellow}
\altaffiltext{\nrao}{National Radio Astronomy Observatory, P.O. Box O, Socorro, NM, 87801}
\altaffiltext{\dark}{Dark Cosmology Centre, Niels Bohr Institute, Copenhagen, Denmark}
\altaffiltext{\edinburgh}{Scottish Universities Physics Alliance, Institute for Astronomy, University of Edinburgh, Royal Observatory, Edinburgh, EH9 3HJ, UK}
\altaffiltext{\gsfc}{NASA/Goddard Space Flight Center, Greenbelt, MD 20771}
\altaffiltext{\jssi}{Joint Space Science Institute, University of Maryland, College Park, MD 20742}
\altaffiltext{\iceland}{Centre for Astrophysics and Cosmology, Science Institute, University of Iceland, Dunhagi 5, 107 Reykjav\'ik, Iceland}
\altaffiltext{\eso}{European Southern Observatory, Alonso de C\'ordova 3107, Vitacura, Casilla 19001, Santiago 19, Chile }
\altaffiltext{\warwick}{Department of Physics, University of Warwick, Coventry CV4 7AL, UK}
\altaffiltext{\leicester}{Department of Physics and Astronomy, University of Leicester, Leicester LE1 7RH, UK}
\altaffiltext{\mail}{e-mail: dperley@astro.caltech.edu .}

\slugcomment{Submitted to ApJ 2014-07-16, accepted 2015-01-12}

\begin{abstract}
Luminous infrared galaxies and submillimeter galaxies contribute significantly to stellar mass assembly and provide an important test of the connection between the gamma-ray burst rate and that of overall cosmic star-formation.   We present sensitive 3 GHz radio observations using the Karl G. Jansky Very Large Array of 32 uniformly-selected GRB host galaxies spanning a redshift range from $0 < z < 2.5$, providing the first fully dust- and sample-unbiased measurement of the fraction of GRBs originating from the Universe's most bolometrically luminous galaxies.  Four galaxies are detected, with inferred radio star-formation rates ranging between 50--300 $M_\odot$yr$^{-1}$.  Three of the four detections correspond to events consistent with being optically-obscured ``dark'' bursts.   Our overall detection fraction implies that between 9--23\% of GRBs between $0.5<z<2.5$ occur in galaxies with $S_{\rm 3 GHz} > 10 \mu$Jy, corresponding to SFR $>$ 50 $M_\odot$yr$^{-1}$ at $z\sim1$ or  $>$ 250 $M_\odot$yr$^{-1}$ at $z\sim2$.  Similar galaxies contribute approximately 10--30\% of all cosmic star-formation, so our results are consistent with a GRB rate which is not strongly biased with respect to the total star-formation rate of a galaxy.  However, all four radio-detected hosts have stellar masses significantly lower than IR/submillimeter-selected field galaxies of similar luminosities.  We suggest that the GRB rate may be suppressed in metal-rich environments but independently enhanced in intense starbursts, producing a strong efficiency dependence on mass but little net dependence on bulk galaxy star-formation rate.
\end{abstract}

\section{Introduction}
\label{sec:intro}

One of the primary appeals of the study of long-duration gamma-ray bursts (GRBs) lies in their potential to address broader questions of cosmology and galaxy evolution.  As the explosions of massive stars at cosmological distances, GRBs are intimately connected with cosmic star-formation, and the evolution of the cosmic GRB rate and the changing demographics of their host galaxies with time should reflect overall cosmological trends and inform our understanding of how galaxies and the stars within them form and evolve over the Universe's history \citepeg{Hogg+1999,Blain+2000,Porciani+2001,Firmani+2004,Natarajan+2005,Kistler+2008,Butler+2010,Tanvir+2012,Robertson+2012,Salvaterra+2013,Trenti+2015}.

Central to the utility of GRBs for this purpose is their extreme luminosity at all electromagnetic wavelengths, including wavelengths unaffected by absorption due to dust and gas.  In particular, GRBs are first detected in hard X-rays and gamma-rays and are now routinely localized to $\sim$2\arcsec\ precision in soft X-rays using the \emph{Swift} X-ray Telescope (\citealt{Burrows+2005}), meaning that---with sufficient follow-up effort---their host-galaxy population and redshift distribution can be constructed independent of the effects of obscuration \citepeg{Hjorth+2012,Jakobsson+2012,Perley+2013a}.  

For this reason, one particular question in galaxy evolution which GRBs showed significant promise to help resolve is the relative importance of extremely luminous, dusty galaxies to cosmic star-formation \citepeg{Djorgovski+2001,RamirezRuiz+2002,Berger+2003}.   The UV/optical tracers by which galaxy and cosmic star-formation rates are normally estimated are significantly impacted by interstellar dust.   Most galaxies are predominately optically thin and the effects of dust can be corrected for via measurements of the UV spectral slope or Balmer decrement \citep{Meurer+1999,Smolcic+2009}.  However, the most bolometrically luminous galaxies (such as submillimeter galaxies [SMGs] and local ultra-luminous infrared galaxies [ULIRGs]) contain sufficient dust so as to be optically thick at UV and optical wavelengths, causing optical/UV-based tracers to inevitably underpredict the true star-formation rates of these galaxies even after dust correction \citep{Blain+2002,Goldader+2002,Chapman+2005}.  Instead, the star-formation rates of these objects (and therefore their contribution to the cosmic SFR) must be measured using long-wavelength tracers such as far-IR or radio continuum, but these methods have historically suffered from limited sensitivity, uncertain calibration, uncertain dust temperatures, and AGN contamination.  Estimates of the importance of the most luminous and dusty galaxies to star-formation have varied widely between different studies---in part due simply to varying definitions of what distinguishes a ``luminous'' and dusty galaxy from an ``ordinary'' one---but have ranged from estimates of 50\% or more \citep{PerezGonzalez+2005,Michalowski+2010,Magnelli+2013} down to only about 10\% \citep{Rodighiero+2011}; see \cite{Casey+2014} for a review.

In principle, the contribution of dusty star-forming galaxies (or any other galaxy population) to overall cosmic star-formation could be determined simply by measuring the fraction of GRBs hosted within such galaxies:  for example, if ULIRGs (galaxies with $L_{\rm IR} > 10^{12} L_\odot$) represent 40\% of all star-formation at $1<z<2$ they should also produce 40\% of all GRBs over the same redshift interval.   However, this would strictly apply only if GRBs represent an unbiased tracer of overall star-formation.   A large volume of evidence now suggests they do not:  GRB host galaxies at $z\lesssim1.5$ have lower masses, lower metallicities, bluer colors, and more irregular morphologies compared to what would be expected from an unbiased SFR-tracing population or when compared to the core-collapse supernova host population \citep{Fruchter+2006,Modjaz+2008,Levesque+2010,Graham+2013,Kelly+2014,Vergani+2014}, even when including dust-obscured ``dark'' bursts (\citealt{Perley+2013a}, although c.f.\ \citealt{Hunt+2014}).

Nevertheless, measuring the contribution of very luminous, dusty galaxies to the GRB rate remains important for understanding the overall link between GRBs and cosmic star-formation, and the influence of dust on our view of each.   What causes GRBs to favor certain environments over others is not well-understood.  A strong preference for low metallicity has been a long-favored explanation given the connection between line-driven winds and angular momentum loss \citep{MacFadyen+1999,Woosley+1999,Woosley+2006}, but a variable IMF or other environmental factors (such as dynamical interaction of stars in dense stellar clusters which preferentially form in the most intense starbursting galaxies; \citealt{vandenHeuvel+2013}) may also help explain the observed characteristics of the host galaxy population.  As extrema of the star-forming galaxy population, dusty and luminous starbursts form an excellent test-bed for distinguishing different hypotheses: in particular, they are probably very metal-rich in their interiors \citep{Nagao+2012} but also very dense and intensely star-forming \citep{Daddi+2007}. 

A number of previous efforts to observe GRB hosts at mid-IR, submillimeter/and or radio wavelengths have been performed over the past decade \citep{Berger+2003,Barnard+2003,Tanvir+2004,LeFloch+2005,Stanway+2010,Hatsukade+2012,Michalowski+2012b,Wang+2012,Perley+2013b,Hunt+2014,Schady+2014}.  However, all of these efforts have been either limited in scope (observing only small samples), limited in sensitivity (insensitive to even very luminous galaxies beyond $z>1$), or subject to uncertain selection biases (most commonly a bias in favor of GRBs with afterglow-based redshift determinations that will systematically miss dust-obscured events, although other efforts have specifically targeted only the \emph{most} heavily dust-obscured GRBs).

In this paper, we present results of the first GRB host galaxy survey that is simultaneously unbiased in regards to target selection, sensitive enough to detect the long-wavelength emission from luminous star-forming galaxies even out to $z\sim2.5$, and large enough to usefully statistically constrain the fraction of GRBs that originate in such systems.  Specifically, we survey a sample of 32 uniformly-selected GRB host galaxies (a factor of $\sim$ 2 larger than any previous long-wavelength survey) to a limiting radio flux density of approximately 10$\mu$Jy at 3 GHz (a factor of 2--3 deeper than any other radio host survey)---detecting four in total and providing the first definitive measurement of the fraction of GRBs produced by the Universe's most luminous galaxies.  Our sample selection, observations, and analysis are discussed in \S \ref{sec:observations}.  We present our results and interpret our four detections as star-formation-associated emission from the GRB host galaxies in \S \ref{sec:results}.  Further discussion of individual systems, including a detailed discussion of all detections and a few notable nondetections, is presented in \S \ref{sec:detections} and \S \ref{sec:nondetections}.   The overall statistical properties of our sample and its implications are discussed in \S \ref{sec:demographics}, and conclusions are summarized in \S \ref{sec:conclusions}.

\section{Sample and Observations}
\label{sec:observations}

\subsection{Sample Selection}

Given the heterogeneous nature of previous long-wavelength efforts and the need to determine the fraction of luminous hosts even among both obscured and unobscured GRBs (a few dusty and luminous galaxies have been reported hosting even the latter), the primary consideration guiding the choice of targets was the need for a uniformly-constructed sample.  The production of unbiased subsets out of what is now a very large (but, often, poorly and non-uniformly observed) overall GRB catalog using simple observability cuts has become increasingly widespread in recent years, starting with the efforts of \cite{Jakobsson+2006} and \cite{Fynbo+2009} to select subsets of GRB afterglows for population analysis and, more recently, the host-galaxy focused TOUGH (The Optically Unbiased GRB Host; \citealt{Hjorth+2012}) and BAT6 \citep{Salvaterra+2012} samples.

For this effort we chose a subset of GRBs from TOUGH, which was well-suited to our goals for several reasons.  First, the TOUGH sample size is large enough to be statistically informative but small enough to be feasible in a one-year Very Large Array (VLA) campaign.  Second, deep optical host-galaxy imaging was available for all targets, needed in order to calculate accurate centroids and angular size constraints.  Third, the survey boasts relatively high redshift completeness ($\sim$85\%, provided from a combination of afterglow spectroscopy available for most events and an extensive host-galaxy spectroscopic campaign; \citealt{Jakobsson+2012} and \citealt{Kruehler+2012}).   Finally, all GRBs in the sample date to 2007 or earlier, ensuring a long temporal baseline between explosion and the present time such that any contribution from a radio afterglow is minimal.

The original selection criteria by which the TOUGH sample was chosen out of the broader catalog of \emph{Swift} GRBs are described in detail in \cite{Hjorth+2012}, which together establish a list of 69 GRBs.  We added two additional critieria for our VLA observations.  First, to ensure observability from the VLA we required a declination of $\delta > -30\arcdeg$ (since the original TOUGH criteria establish a declination maximum of $\delta < +27\arcdeg$, this effectively limits the sample to $-30\arcdeg < \delta < 27\arcdeg$.)   In addition, we excluded any targets known to be at $z>2.5$, since detecting even very luminous galaxies at higher redshift is extremely difficult at radio wavelengths.  We did, however, include TOUGH GRBs in our declination range whose redshift is not yet known, to ensure that imposing a redshift cut did not produce any biases (by, for example, inadvertently excluding dust-obscured events at $z<2.5$ whose redshifts may have been missed).

These additional cuts on declination and redshift produced a final sample of 32 GRBs.  Of these, 27 are known to be at $z<2.5$ (although one of these does not have a unique redshift/host identification) and 5 are at unknown redshift.  

\begin{deluxetable*}{llll llr ll}  
\tabletypesize{\footnotesize}
\tablecaption{VLA Observations}
\tablecolumns{9}
\tablehead{
\colhead{GRB} &
\colhead{RA\tablenotemark{a}} & 
\colhead{Dec\tablenotemark{a}} & 
\colhead{$z$\tablenotemark{b}} & 
\colhead{Config.\tablenotemark{c}} &
\colhead{Observation dates} &
\colhead{$t_{\rm int}$\tablenotemark{d}} &
\colhead{Beam size\tablenotemark{e}} &
\colhead{RMS noise\tablenotemark{f}}
 \\
\colhead{} &
\colhead{} &
\colhead{} &
\colhead{} &
\colhead{} &
\colhead{(UT)} &
\colhead{(min)} &
\colhead{($\arcsec$)} &
\colhead{($\mu$Jy/beam)}
}
\startdata
 050416A & 12:33:54.64 &  +21:03:26.8 & 0.654 & A    & 2014-01-02             & 119.7 & 2.3$\times$2.1 & 5.6 \\
 050525A & 18:32:32.67 &  +26:20:21.6 & 0.606 & A,B  & 2012-12-02, 2014-01-10 & 138.7 & 1.9$\times$1.6 & 3.0 \\
 050714B & 11:18:47.71 &$-$15:32:49.0 & 2.438 & BnA  & 2014-01-27             & 122.9 & 2.4$\times$1.7 & 3.3 \\ 
 050801  & 13:36:35.32 &$-$21:55:42.7 & 1.560 & BnA  & 2014-01-30             & 118.1 & 1.9$\times$1.6 & 3.3 \\ 
 050803  & 23:22:37.85 &  +05:47:08.5 & ?     & A    & 2012-12-24, 2013-01-05 & 141.5 & 0.7$\times$0.6 & 3.5 \\
 050819  & 23:55:01.62 &  +24:51:39.0 & 2.500 & B    & 2013-11-01             & 121.3 & 2.5$\times$2.1 & 3.6 \\
 050824  & 00:48:56.21 &  +22:36:33.1 & 0.830 & B    & 2013-11-23             & 119.7 & 2.2$\times$2.0 & 3.5 \\
 050922B & 00:23:13.37 &$-$05:36:16.7 & ?     & A,B\tablenotemark{g}  & 2012-12-03, 2013-11-22 & 161.1 & 0.8$\times$0.6 & 7.8 \\  
 050922C & 21:09:33.08 &$-$08:45:30.2 & 2.198 & A,B  & 2012-10-08, 2013-11-17 & 145.5 & 2.3$\times$1.6 & 3.4 \\
 051006  & 07:23:14.14 &  +09:30:20.0 & 1.059 & A,B  & 2013-01-05, 2013-12-07 & 134.0 & 1.7$\times$1.4 & 3.2 \\
 051016B & 08:48:27.85 &  +13:39:20.4 & 0.936 & B    & 2013-12-06             & 118.1 & 2.6$\times$2.0 & 3.9 \\
 051117B & 05:40:43.38 &$-$19:16:27.2 & 0.481 & A,BnA& 2012-11-01, 2014-01-27 & 140.6 & 1.5$\times$1.2 & 5.4 \\
 060218  & 03:21:39.69 &  +16:52:01.6 & 0.033 & A    & 2012-11-30             & 74.5  & 0.6$\times$0.6 & 3.9 \\ 
 060306  & 02:44:22.88 &$-$02:08:54.7 & 1.55  & A    & 2012-12-15             & 70.7  & 0.7$\times$0.5 & 6.2 \\ 
 060604  & 22:28:55.04 &$-$10:54:56.1 & 2.140 & A,B  & 2012-12-30, 2013-11-23 & 137.8 & 2.3$\times$1.5 & 3.2 \\
 060805A & 14:43:43.47 &  +12:35:11.2 & 0.60/2.44 & A & 2012-11-11         & 109.8 & 0.7$\times$0.6 & 3.3 \\
 060814  & 14:45:21.36 &  +20:35:09.2 & 1.920 & B    & 2013-11-17             & 119.7 & 2.2$\times$2.0 & 3.2 \\
 060908  & 02:07:18.42 &  +00:20:32.2 & 1.884 & B    & 2013-12-10             & 121.3 & 2.2$\times$1.9 & 9.5 \\
 060912A & 00:21:08.13 &  +20:58:18.5 & 0.937 & A,B  & 2013-01-01, 2013-11-03 & 134.0 & 1.8$\times$1.6 & 3.1 \\
 060923A & 16:58:28.14 &  +12:21:37.9 & 2.5   & B    & 2013-11-23             & 118.1 & 2.3$\times$2.1 & 6.1 \\
 060923C & 23:04:28.36 &  +03:55:28.4 & ?     & A,B  & 2012-10-25, 2013-11-22 & 160.5 & 1.9$\times$1.5 & 3.2 \\
 061021  & 09:40:36.12 &$-$21:57:05.2 & 0.346 & A,BnA& 2012-11-15, 2014-01-29 & 140.6 & 1.5$\times$1.1 & 3.0 \\
 061110A & 22:25:09.89 &$-$02:15:30.4 & 0.758 & A,B  & 2012-12-24, 2013-10-26 & 139.6 & 3.1$\times$1.5 & 6.1 \\ 
 061121  & 09:48:54.59 &$-$13:11:42.1 & 1.314 & B    & 2014-01-02             & 118.1 & 3.0$\times$1.9 & 5.8 \\
 070129  & 02:28:00.98 &  +11:41:03.4 & 2.340 & B    & 2013-12-03             & 119.7 & 2.2$\times$1.9 & 5.1 \\
 070224  & 11:56:06.57 &$-$13:19:48.8 & 1.992 & B    & 2013-11-23             & 118.1 & 3.0$\times$1.9 & 5.7 \\ 
 070306  & 09:52:23.29 &  +10:28:55.5 & 1.496 & A    & 2012-11-11             & 189.5 & 0.7$\times$0.6 & 2.9 \\
 070506  & 23:08:52.31 &  +10:43:20.8 & 2.310 & A,B  & 2013-01-01, 2013-11-03 & 132.2 & 1.9$\times$1.6 & 5.4 \\
 070611  & 00:07:58.12 &$-$29:45:20.4 & 2.040 & A,BnA& 2012-11-08, 2014-02-04 & 140.6 & 1.7$\times$1.5 & 3.1 \\
 070621  & 21:35:10.08 &$-$24:49:03.1 & ?     & A    & 2012-10-31             & 127.9 & 1.0$\times$0.6 & 6.2 \\
 070808  & 00:27:03.36 &  +01:10:34.4 & ?     & B    & 2013-11-23             & 122.9 & 2.6$\times$2.0 & 3.4 \\
 070810A\tablenotemark{h} & 12:39:51.24 &  +10:45:03.2 & 2.170 & A    & 2012-11-11             & 69.8  & 0.7$\times$0.6 & 4.6 \\  
\enddata 
\tablenotetext{a}{\ Observation pointing center (J2000).}
\tablenotetext{b}{\ Redshift of host or afterglow, generally from \cite{Hjorth+2012} and sources quoted within or from \citealt{Kruehler+2015}.  GRB 060805A has two possible host-galaxy candidates; we will generally assume $z=2.44$ in our plots since it is closer to the Swift median redshift.  The redshifts of GRB 060306 and GRB 060923A are from \cite{Perley+2013a}; the redshift of GRB 060923A is photometric.}
\tablenotetext{c}{\ VLA array configuration.}
\tablenotetext{d}{\ Total time on-source, excluding overheads.}
\tablenotetext{e}{\ Major and minor axis FWHM of the synthesized beam.}
\tablenotetext{f}{\ Noise (1$\sigma$) estimated from the standard deviation of 1000 randomly chosen points in the final map.}
\tablenotetext{g}{\ Only the A-configuration observations were usable due to severe RFI affecting the B-configuration dataset.}
\tablenotetext{h}{\ GRB 070810A was observed as part of the TOUGH campaign although it technically did not satisfy the sun-angle constraint of the survey.  We include it as part of our survey here.}
\label{tab:observations}
\end{deluxetable*}

\subsection{VLA Observations}

All of the 32 targets described above were observed with the fully-upgraded Karl G. Jansky Very Large Array, using the S-band receivers (central frequency 3 GHz) and 8-bit samplers with a bandwidth of 2048 MHz.   Observations were conducted during the A-configuration cycle of fall 2012, the B-configuration cycle of fall 2013, and the BnA-configuration of winter 2014 (project codes 12B-305 and 13B-316). Integration times for each field were approximately two hours on-target.    Observations employed the WIDAR correlator \citep{PerleyR+2011} using a spectral resolution of 2 MHz and a sampling time of 4 seconds.  Local phase and amplitude calibration was established via observations of a nearby source approximately every six minutes, and the flux density scale was calibrated with a single observation of 3C48, 3C138, or 3C286 using the reference scale of \cite{PerleyButler2013} at the beginning or end of each observation sequence.  A summary of all observations is presented in Table \ref{tab:observations}.

Phase stability was very good, with typical variations of less than 10 degrees over the 6-minute calibration cycle.  More troublesome were the receiver amplitude gain variations induced by transmission from geostationary satellites. As seen from the VLA's latitude, these satellites are located in a belt near a declination of $-$4.5\arcdeg, and examination of our data shows that observations of sources located within about 10 degrees of this declination (i.e., +5.5\arcdeg\ to $-$14.5\arcdeg) show significant gain variations due to variable power from these satellites entering the analog signal path through the antennas' sidelobes.  As the antennas track the source, the input satellite power varies as the sidelobes sweep past the satellites. These gain variations are tracked by the antennas' on-board switched power calibration system, and application of this monitoring system reduced the gain uncertainty to less than 5\% for all affected sources.  For those sources within $\sim$2--3 degrees of declination $-$4.5\arcdeg\ (050922B, 060306, 061110A) the satellite input power occasionally saturated the analog receiver system.  The data from those times were deleted.

\begin{figure*}
\centerline{
\includegraphics[width=6.9in,angle=0]{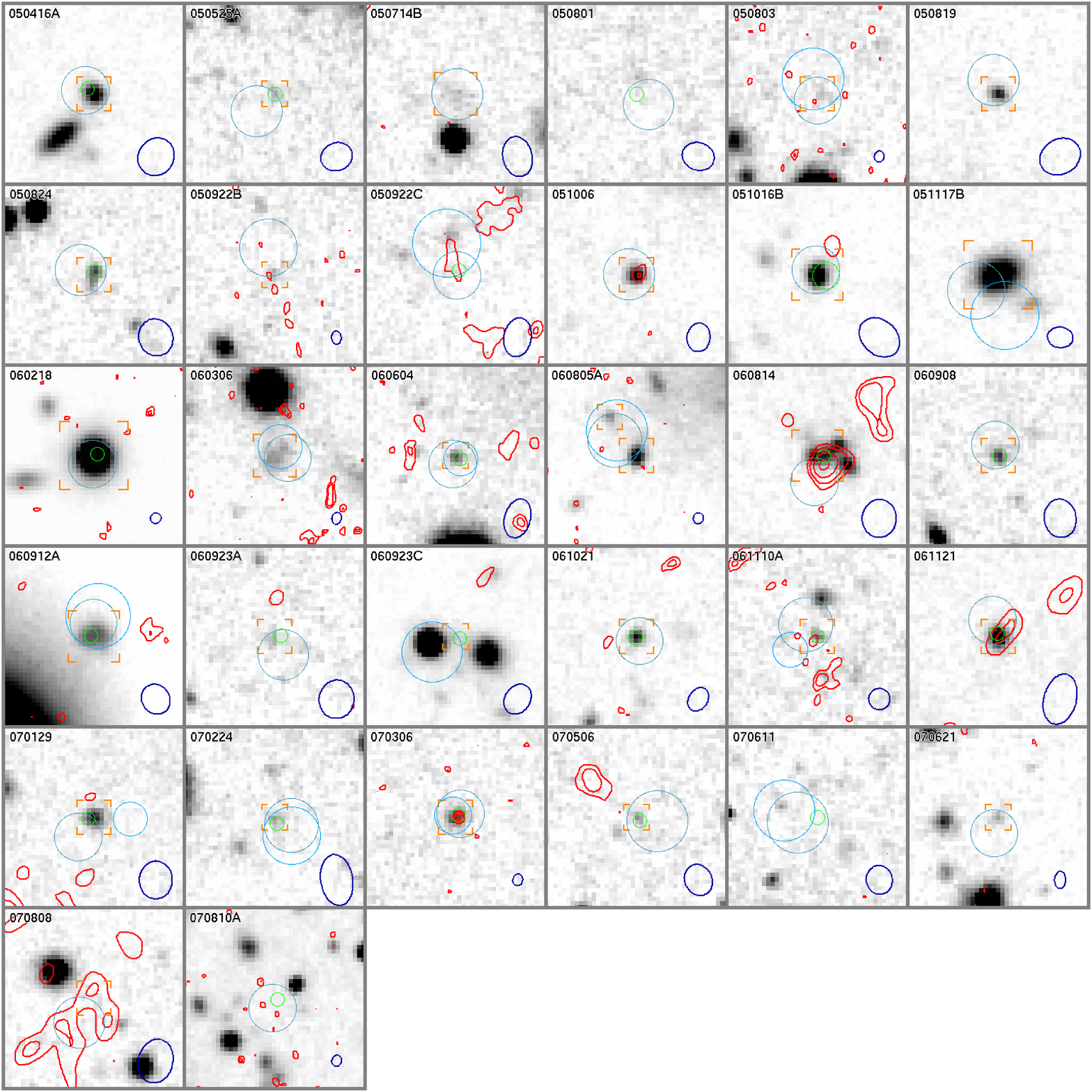}} 
\caption{Mosaic of TOUGH $R$-band imaging of 32 GRB host galaxy positions, with VLA 3 GHz flux-density contours overlaid in red.  Contour levels are set to 2.3, 3.0, 4.0, 5.0, and 6.0 times the VLA image RMS.  (Several images have no flux above the 2.3$\sigma$ contour within the 10.5\arcsec$\times$10.5\arcsec\ thumbnail.)  The GRB afterglow location is represented by light blue circles (XRT positions from \citealt{Butler2007} or \citealt{Evans+2009}) or by a green circle (optical/IR positions).  The host galaxy, when detected in the optical images, is centered in the image and identified by an orange box reticle.  The VLA synthesized beam is indicated by a dark blue ellipse at bottom right.}
\label{fig:images}
\end{figure*}

Data reduction was carried out using the Astronomical Image Processing System (AIPS).   Following gain and amplitude calibration, the data were examined for sporadic radio frequency interference (RFI).  Visibilities with amplitude values on the (2 MHz, 4 sec) resolution scale exceeding 5-sigma were removed.  In addition, the data within the transmission bands for the XM/Sirius satellite digital radio systems (2320 to 2350 MHz) were removed for all observations. Imaging/deconvolution was done using faceting over the primary beam to reduce non-coplanar distortions.  For most targets, no attempt was made to account for the spectral index gradients of the background sources due to the wavelength-dependent primary beam size, as these background sources were generally far enough from the target position that the imaging errors were reduced to below the noise level at the position of the target.  One exception is the field of GRB 061110A, which contains several strong sources; for this target we imaged each frequency window individually and averaged together the resulting maps.  The other exception was for the field of GRB 070808, in which the target position was located along an image artifact associated with a single moderately-strong quasar.  Effective RMS sensitivities of the final maps are between 3--4~$\mu$Jy/beam for most targets.  A few fields are slightly noisier (4--9 $\mu$Jy), in most cases due to the influence of geostationary-belt satellites.  A mosaic of all fields is shown in Figure \ref{fig:images}.

\subsection{VLA Flux Density Measurement}

We evaluate the significance of host detections/limits, and measure the host-galaxy fluxes, in two different ways.  First, we take the position of the host centroid in the optical images and simply measure the flux-density value in the 0.18\arcsec\ cell nearest to this position in the reduced, deconvolved VLA map.  To measure the uncertainty we take the RMS of the same final map, evaluated directly by a Monte-Carlo sampling of the fluxes of 1000 randomly-selected cells in the central region of the image.  The positions and fluxes evaluated using this method are presented in Table \ref{tab:fluxes}.

There is a possibility that the radio host centroid may be offset from the optical one, and there are a few fields where the exact host position is not known because the host is not detected in the available TOUGH imaging.  To accommodate the possibility of an offset host galaxy, we also employ a second method in which we take the maximum value of any cell within a radius given by either the measured size of the host galaxy in the Very Large Telescope (VLT) optical images or (if not detected) by the size of the positional uncertainty region of the afterglow in the \emph{Swift} XRT image.  Uncertainties are estimated by a similar 1000 Monte-Carlo sampling as for the fixed-position.  Fluxes calculated via this method were (for all secure detections) consistent within uncertainties with the fixed-position values, and we conservatively adopt the lower fluxes of the fixed-position method for our later analysis.   However, we do use the maximum-flux method to check the confidence of our detections:  specifically, we consider as a ``detection'' any source with a measured fixed-position flux at least 2.5$\sigma$ times the uncertainty and for which the flux measured from the maximum-flux method exceeds the flux measured in 98\% of the maximum-flux Monte Carlo trials (rightmost column in Table \ref{tab:fluxes}).  We also verify that any detections are pointlike (see \S \ref{sec:results}).

Our flux measurement procedure assumes that the host galaxy is not extended at the resolution of our available imaging.  This is generally a good assumption (most hosts are only marginally resolved even at the resolution of the VLT images, and the largest galaxies are only $\sim$2$\arcsec$ in diameter, comparable to our beam size in most cases), although for a few fields with more extended host galaxies and/or compact beams (e.g., 060814, 060218, 051117B) our flux measurement/limit may be a modest underestimate of the true integrated value.

\begin{deluxetable}{lrrl}  
\tabletypesize{\small}
\tablecaption{Host Galaxy 3 GHz Flux Density Measurements}
\tablecolumns{4}
\tablehead{
\colhead{GRB} &
\colhead{$F_\nu$\tablenotemark{a}} &
\colhead{$F_\nu^{max}$ \tablenotemark{b}} &
\colhead{Conf.\tablenotemark{c}}
\\
\colhead{} &
\colhead{($\mu$Jy)} &
\colhead{($\mu$Jy)} &
\colhead{}
}
\startdata
050416A &$-$7.87 $\pm$ 4.18  &  -0.64 &  0.150 \\
050525A &$-$1.25 $\pm$ 2.99  &   0.57 &  0.221 \\
050714B &   4.91 $\pm$ 3.17  &   6.48 &  0.799 \\ 
050801  &$-$4.79 $\pm$ 3.64  &   1.79 &  0.103 \\
050803  &   1.46 $\pm$ 3.42  &   8.78 &  0.756 \\
050819  &$-$0.92 $\pm$ 3.22  &   1.72 &  0.329 \\
050824  &$-$0.11 $\pm$ 3.51  &   0.70 &  0.147 \\
050922B &$-$2.72 $\pm$ 8.01  &  22.70 &  0.965 \\
050922C &   8.75 $\pm$ 3.54  &   9.37 &  0.973 \\
\textbf{051006}  &   9.08 $\pm$ 3.17  &   9.74 &  0.984 \\
051016B &   0.97 $\pm$ 4.03  &   8.25 &  0.874 \\
051117B &$-$4.87 $\pm$ 4.49  &   5.15 &  0.221 \\
060218  &   5.52 $\pm$ 3.88  &  10.38 &  0.303 \\
060306  &   7.03 $\pm$ 6.40  &   8.56 &  0.079 \\
060604  &$-$3.50 $\pm$ 3.35  &   1.17 &  0.266 \\
060805A\tablenotemark{d} &$-$4.52 $\pm$ 3.45  &   8.43 &  0.619 \\ 
060805A\tablenotemark{e} &   2.25 $\pm$ 3.45  &   5.11 &  0.426 \\  
\textbf{060814}  &  11.34 $\pm$ 3.12  &  15.32 &  0.998 \\
060908  &   4.53 $\pm$ 5.95  &   8.04 &  0.692 \\
060912A &   4.54 $\pm$ 3.37  &   6.87 &  0.850 \\
060923A &   0.05 $\pm$ 5.00  &   5.45 &  0.685 \\
060923C &   2.87 $\pm$ 3.89  &   5.95 &  0.736 \\
061021  &   0.82 $\pm$ 2.99  &   4.41 &  0.563 \\
061110A &  14.20 $\pm$ 6.08  &  18.64 &  0.978 \\
\textbf{061121}  &  17.07 $\pm$ 5.47  &  17.80 &  0.995 \\
070129  &   4.92 $\pm$ 5.23  &  12.22 &  0.918 \\
070224  &$-$1.35 $\pm$ 6.75  &   1.53 &  0.335 \\
\textbf{070306}  &  11.31 $\pm$ 2.84  &  13.31 & 0.998 \\
070506  &   3.69 $\pm$ 4.62  &   6.74 &  0.750 \\
070611  &$-$9.62 $\pm$ 4.53  &   3.74 &  0.086 \\
070621  &  11.35 $\pm$ 6.45  &  13.19 &  0.696 \\
070808  &   9.78 $\pm$ 3.47  &  10.63 &  0.971 \\
070810A &$-$1.45 $\pm$ 4.76  &   8.33 &  0.480 \\
\enddata
\label{tab:fluxes}
\tablenotetext{a}{\ Measured flux density (at 3 GHz) at the position of the optical/NIR host galaxy centroid (or best-position afterglow centroid, if the host is not detected in optical/NIR observations.)}
\tablenotetext{b}{\ Maximum flux density (at 3 GHz) in any 0.18\arcsec\ synthesized cell consistent with the position of the optical/NIR host galaxy disk (or afterglow uncertainty region, if the host is undetected).}
\tablenotetext{c}{\ Significance of the detection, based on placing a large number of search apertures of identical size to the aperture used to calculate $F_\nu^{max}$ randomly across the image and calculating the maximum flux density in each one.}
\tablenotetext{d}{\ Southern host candidate ($z=0.60$).}
\tablenotetext{e}{\ Northern host candidate ($z=2.44$).}
\end{deluxetable}

\subsection{Optical/IR Observations}

As discussed in the next section, four of our host-galaxy targets were detected (or are very likely detected).  To investigate these systems in more detail, we also acquired deep optical, NIR, and mid-IR imaging from a variety of sources in order to perform SED modeling and measure masses and UV-based star-formation rates.  Many of these data points were taken from the literature (in particular \citealt{Hjorth+2012}, \citealt{Kruehler+2011}, and \citealt{Perley+2013a}), including all data points for the host galaxies of GRB 060814 and GRB 070306.  New observations are briefly summarized below.

GRBs 051006 and 061121 were observed with the Low-Resolution Imaging Spectrometer (LRIS; \citealt{Oke+1995}) at Keck Observatory on a number of occasions using several different filters between 2005 and 2014.  Observations were reduced using the custom IDL pipeline \texttt{lpipe}\footnote{\url{http://www.astro.caltech.edu/~dperley/programs/lpipe.html}}.  Neither of these fields overlaps with the Sloan Digital Sky Survey (SDSS) footprint, so secondary standards within each field were calibrated by acquiring observations of each field with the Palomar 60-inch robotic telescope (P60; \citealt{Cenko+2006}) in the $uBgriz$ filters and calibrating relative to \cite{Landolt+2009} standards and to various SDSS fields observed the same night during photometric weather.   Magnitudes of each galaxy are measured in the reduced images using standard aperture photometry techniques.

GRB 061121 was also observed with the Wide Infrared Camera (WIRC) at Palomar Observatory in $J$-band using the replacement engineering-grade detector on 2014-05-10 UT; we acquired 22 60-second exposures on the field.   GRB 051006 was observed using the same instrument on 2014-10-11 UT (we acquired 72 15-second exposures in both $J$-band and $K_s$ bands, although only 42 exposures were included in the final stack due to suboptimal dithering) and on 2014-10-19 UT (we acquired 29 40-second exposures in $H$-band) in all three broad-band NIR filters.  Data were reduced using custom IDL scripts.  Magnitudes of the host galaxy were measured using aperture photometry with the calibration established via 2MASS stars in the field.

GRBs 051006 and 061121 were observed by the Infrared Array Camera (IRAC) on board the the Spitzer Space Telescope during Cycle 9 (GO 90062, PI D. Perley).  We used the PBCD reduced files from the \emph{Spitzer} legacy archive and measure magnitudes using the standard zeropoints in the Spitzer handbook using an aperture radius of 1.2\arcsec.

All host-galaxy photometry for these two objects is presented in Table \ref{tab:optical} and in Figure \ref{fig:multised}.

\begin{deluxetable}{llrl}  
\tabletypesize{\small}
\tablecaption{Host Galaxy Optical/NIR Photometry}
\tablecolumns{4}
\tablehead{
\colhead{Filter} &
\colhead{Mag\tablenotemark{a}} &
\colhead{Flux\tablenotemark{b}} &
\colhead{Instrument} \\
\colhead{} &
\colhead{} &
\colhead{($\mu$Jy)} &
\colhead{}}
\startdata
\multicolumn{4}{c}{GRB 051006} \\
$u$     &  24.50 $\pm$ 0.20 &   0.75 $\pm$ 0.15 & Keck-I/LRIS \\
$B$     &  24.03 $\pm$ 0.10 &   1.31 $\pm$ 0.13 & Keck-I/LRIS \\
$V$     &  23.44 $\pm$ 0.10 &   1.89 $\pm$ 0.18 & Keck-I/LRIS \\ 
$R$     &  22.99 $\pm$ 0.07 &   2.33 $\pm$ 0.15 & VLT-U1/FORS2\tablenotemark{c} \\
$i$     &  22.51 $\pm$ 0.15 &   4.09 $\pm$ 0.61 & Keck-I/LRIS \\
$z$     &  22.08 $\pm$ 0.10 &   6.00 $\pm$ 0.58 & Keck-I/LRIS \\
$J$     &  20.76 $\pm$ 0.18 &   8.37 $\pm$ 1.51 & P200/WIRC \\ 
$H$     &  19.82 $\pm$ 0.15 &  12.52 $\pm$ 1.85 & P200/WIRC \\
$K_s$   &  19.04 $\pm$ 0.16 &  16.53 $\pm$ 2.62 & P200/WIRC \\
$3.6$   &  17.72 $\pm$ 0.05 &  22.91 $\pm$ 1.08 & Spitzer/IRAC\\
\hline
\multicolumn{4}{c}{GRB 061121} \\
$u$     &  23.08 $\pm$ 0.10 &   3.98 $\pm$ 0.38 & Keck-I/LRIS \\
$B$     &  23.34 $\pm$ 0.05 &   3.33 $\pm$ 0.16 & Keck-I/LRIS \\ 
$g$     &  22.95 $\pm$ 0.04 &   3.95 $\pm$ 0.15 & Keck-I/LRIS \\
$R$     &  22.75 $\pm$ 0.04 &   3.49 $\pm$ 0.13 & VLT-U1/FORS2\tablenotemark{c} \\
$i$     &  22.66 $\pm$ 0.10 &   4.12 $\pm$ 0.40 & Keck-I/LRIS \\
$z$     &  22.33 $\pm$ 0.08 &   5.29 $\pm$ 0.40 & Keck-I/LRIS \\
$J$     &  21.29 $\pm$ 0.28 &   5.49 $\pm$ 1.61 & P200/WIRC   \\
$K$     &  20.14 $\pm$ 0.19 &   6.16 $\pm$ 1.18 & VLT-U1/ISAAC\tablenotemark{c}\\
$3.6$   &  18.72 $\pm$ 0.10 &   9.12 $\pm$ 0.88 & Spitzer/IRAC\\
\enddata
\tablenotetext{a}{Apparent magnitudes of the host galaxy, not corrected for foreground extinction.  SDSS filters (lowercase) are reported in the SDSS sytem (nearly AB; \citealt{Fukugita+1996}).  Other filters are reported as Vega magnitudes.}
\tablenotetext{b}{Host galaxy flux, corrected for foreground extinction.}
\tablenotetext{c}{From \cite{Hjorth+2012}}
\label{tab:optical}
\end{deluxetable}

\section{Results}
\label{sec:results}

A mosaic of our fields is shown in Figure \ref{fig:images}.  The grayscale background image represents the TOUGH $R$-band optical imaging from \cite{Hjorth+2012}; overlaid contours are from our new VLA observations.

The large majority of the hosts are not detected (the flux at or near the GRB location is consistent with the value expected for random locations in the same image; Table \ref{tab:fluxes}), which is no surprise---previous GRB host radio surveys have also produced low detection rates, and only a small fraction of cosmic star-formation beyond $z>0.1$ is in galaxies luminous enough to be individually detected at radio wavelengths, even at these deep levels \citepeg{Karim+2011,Dunne+2009}.

We do, however, clearly detect at least three (and very likely four) of our targets.  Radio sources are detected at the locations of GRBs 060814, 061121, and 070306 at high significance (99\% confidence or greater, based on our Monte-Carlo analysis).  A fourth radio source at the location of GRB 051006 has somewhat lower significance (98\%), but the point-like nature of the object and its near-exact consistency with the optical host localization suggest that it is likely real as well.

A plot summarizing our detections and nondetections (along with a sample of measured or predicted fluxes from a sample of field galaxies; \S \ref{sec:detrate}) is shown in Figure \ref{fig:radioflux}.

\subsection{Star Formation, Afterglow, or AGN?}

While it is impossible to determine with certainty whether or not any individual detection represents emission associated with star-formation in the host galaxy or some other source, it is very likely that all of these detections do indeed correspond to the host-galaxy synchrotron continuum we are seeking.  

There are three possible alternatives: emission from an unrelated foreground/background source, emission from an AGN within the host galaxy, or emission from the GRB afterglow.   The first case can be ruled out on statistical grounds: only about 0.1\% of the sky area in our fields contains a detected source, so the probability of even one chance intersection of a source anywhere in our sample with an unrelated object is low ($\sim$3\%).   An AGN origin within the host is also unlikely, for similar statistical reasons:  most sources with a radio flux density close to 10$\mu$Jy flux are star-forming galaxies, not AGN \citep{Kimball+2011,Condon2012}.  The optical spectroscopy of these systems, where available \citepeg{Jaunsen+2008,Jakobsson+2012}, also shows no evidence for AGN features.

The afterglow possibility is most difficult to generically rule out:  while all of our targets were observed long after the occurrence of the GRB (at least 5 years and typically 7--8 years), few GRBs have been followed to faint enough flux/luminosity limits to directly establish the ``typical''  distribution of afterglow fluxes on multi-year timescales.  The discussion in \cite{Perley+2013b} suggests that fewer than 10\% of all GRB afterglows will be detectable at the flux levels and timescales of our study, but even 10\% contamination would be sufficient to produce most of our detections, given the large overall sample size.   However, it is possible to examine systems individually to rule out this possibility by other means: in particular, by the observation of any of the following: (1) lack of significant fading versus early-time measurements or limits; (2) physical extension of the radio source; or (3) confirmation of a star-forming galaxy-like SED from IR or submillimeter observations.  

GRBs 061121 and 070306 were both observed at radio wavelengths a few days after the GRB and not detected to a flux density limit of $\sim100$ $\mu$Jy \citep{GCN5871,GCN6180}.  The persistence of a 10--20 $\mu$Jy radio afterglow at the time of our observations would require these events to have faded by a factor of only 10 or less between $\sim$3 days and 2000 days, which is exceptionally unlikely given the behavior of all other known radio afterglows, which invariably show steep power-law decays beginning at $\sim$100 days post-GRB or earlier (\citealt{Chandra+2012}).  In the case of GRB 061121, even the possibility of an unusually persistent/late-fading source can be ruled out directly since a second epoch showing clear fading was reported by \cite{GCN5874}.   GRB 070306 also has sensitive FIR observations available (from the \emph{Herschel} survey of \citealt{Hunt+2014}) and the host is well-detected in that data, consistent with its identification as a star-forming galaxy. 

GRBs 060814 and 051006 have no observations available that would enable us to directly rule out an afterglow origin: the sources do not appear obviously extended in our VLA imaging (although the beam size of both observations is comparable to the optical host diameter) and neither was observed in the radio band before our observations. However, given that our optical observations (\S \ref{sec:detections}) indicate star-formation rates consistent with those inferred from our measured radio fluxes it is very likely that these detections represent host-galaxy emission as well.\footnote{Even if these detections were partially due to afterglow, the overall conclusion of this work---that the radio measurements rule out any significant excess star-formation beyond that inferred optically for these objects---would be unaltered.}

\section{Analysis of Detected Sources}
\label{sec:detections}

\begin{deluxetable*}{lllllll}  
\tabletypesize{\small}
\tablecaption{Physical Properties of Radio-Detected Host Galaxies}
\tablecolumns{6}
\tablehead{
\colhead{GRB} &
\colhead{$z$} &
\colhead{$M_*$} &
\colhead{$t_{\rm burst}$\tablenotemark{a}} &
\colhead{$A_V$} &
\colhead{SFR$_{\rm SED}$\tablenotemark{b}} &
\colhead{SFR$_{\rm radio}$\tablenotemark{c}} \\
\colhead{} &
\colhead{} &
\colhead{$M_\odot$} &
\colhead{Myr} &
\colhead{mag} &
\colhead{$M_\odot$yr$^{-1}$} &
\colhead{$M_\odot$yr$^{-1}$}
}
\startdata
 051006  & 1.059 & $1.3_{-0.1}^{+0.1} \times 10^{10}$ & 128 & $1.73_{-0.01}^{+0.03}$ & $  98_{-1}^{+2}   $ & $ 51_{-18}^{+22} $  \\
 060814  & 1.923 & $1.6_{-0.6}^{+1.4} \times 10^{10}$ & 209 & $1.17_{-0.17}^{+0.05}$ & $ 209_{-53}^{+27} $ & $256_{-70}^{+160}$ \\
 061121  & 1.314 & $1.5_{-0.6}^{+0.6} \times 10^{10}$ & 179 & $0.45_{-0.26}^{+0.16}$ & $  27_{-6}^{+27}  $ & $160_{-51}^{+58} $ \\
 070306  & 1.496 & $5.0_{-0.2}^{+0.1} \times 10^{10}$ &  20 & $0.21_{-0.10}^{+0.13}$ & $  17_{-5}^{+7}   $ & $143_{-35}^{+61} $ \\
\enddata
\tablenotetext{a}{Age of the young (starburst) component in the stellar population-synthesis model fit.}
\tablenotetext{b}{Star-formation rate derived from fitting the UV-optical-IR SED of the galaxy.}
\tablenotetext{c}{Star-formation rate derived from the measured radio flux density at the optical host centroid, using the conversion from \cite{Murphy+2011}.}
\label{tab:hostproperties}
\end{deluxetable*}

Since a very high star-formation rate is required for a distant galaxy to be detectable in the radio band, there is good reason to expect that the general properties of the hosts detected in our observations will differ from the overall GRB host population and from ``typical'' star-forming galaxies; comparison of these galaxies to other radio/submillimeter-selected galaxy populations may also provide insight into the nature of the GRB-galaxy connection overall. 

\begin{figure}
\centerline{
\includegraphics[scale=0.52,angle=0]{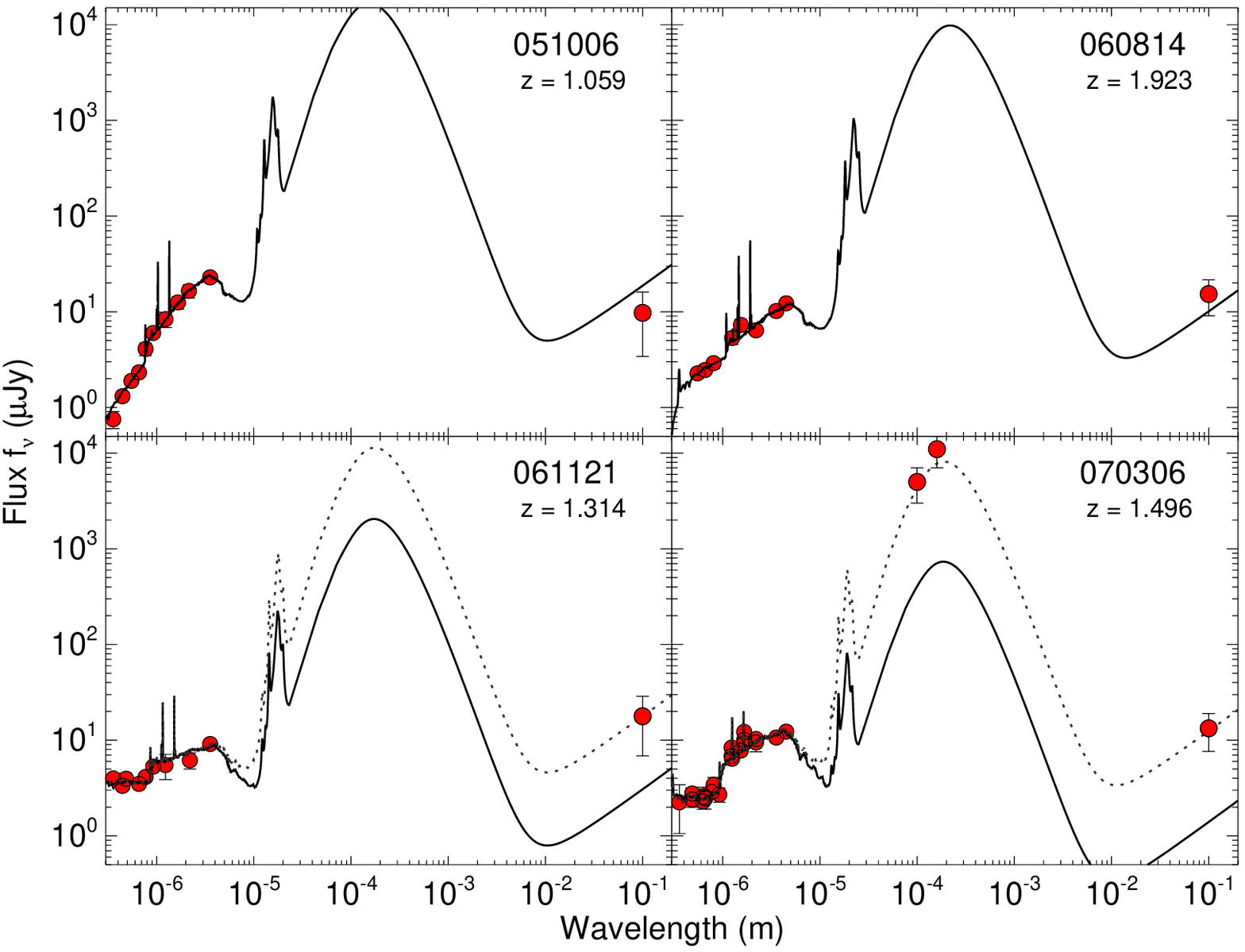}} 
\caption{UV-to-radio SEDs of the host galaxies for all four radio-detected sources in our survey.   The solid curve shows our multiwavelength SED model of each galaxy fit using the UV-optical-NIR data.   In two cases this model accurately predicts the radio flux, indicating that the starburst is optically thin.  For the remaining sources the UV-inferred star-formation alone underpredicts the radio flux, and a heavily obscured nuclear starburst (dashed line) is also needed to explain the optical and radio measurements simultaneously.}
\label{fig:multised}
\end{figure}

Thanks to a combination of photometry from the literature and our own additional observations, we have excellent optical and infrared photometric data for all four radio-detected systems in our sample (although not yet for the nondetections).  After correcting for foreground extinction \citep{Schlegel+1998}, we analyzed the optical-NIR SEDs of these galaxies using a similar procedure as previously employed in \cite{Perley+2013a}, using a custom SED-fitting code based on the stellar population-synthesis templates of \cite{bc03} assuming a \cite{Chabrier2003} IMF and metallicity of 0.5 Solar.\footnote{We also attempted models of 1.0 and 0.2 Solar metallicity and found the derived parameters to be generally consistent with the 0.5 Solar model, but in all cases with significantly higher $\chi^2$ residuals.}   The star-formation history of each galaxy is fit as a two-population model with the maximum age of the older population fixed to the age of the Universe at the redshift of the host, and the maximum age of the younger population free (but required to be at least 20 Myr).  Both populations are assumed to have a continuous star-formation history from the maximum age until today, so the overall star-formation history is constant with an abrupt increase at an arbitrary time $t_{\rm burst}$, the age of the current ongoing starburst.  Results from these fits are shown in Table \ref{tab:hostproperties} and Figure \ref{fig:multised}.

Radio star-formation rates are calculated from our observed VLA fluxes using Equation 17 of \cite{Murphy+2011} and assuming a synchrotron spectral index ($F_\nu \propto \nu^{-\alpha}$) of $\alpha=0.75$; the contribution of free-free radiation is assumed to be negligible at this frequency.  We note that the calibration of radio star-formation rates is subject to some systematic uncertainty at about the level of a factor of $\sim$2, and the use of other relations would produce values offset by a common factor; in particular the measured SFRs and limits would drop by about 40\% if we instead used Equation 15 of \citealt{Yun+2002}.

For two of our galaxies (GRB 051008 and GRB 060814), the star-formation rates inferred from the SED fitting procedure and from our radio observations are consistent; that is, an SED model based only on the optical/NIR photometry accurately predicts (within 2$\sigma$) the observed radio flux.  For the remaining two sources (GRB 070306 and, marginally, GRB 061121), the radio detection is significantly in excess of what would be expected from the SED fit, indicating that an additional, heavily obscured star-forming component is present in these galaxies.  We implement this in Figure \ref{fig:multised} by simply adding an additional component with parameters of $A_V = 15$ mag, $t_{\rm burst} = 10$ Myr, and SFR = SFR$_{\rm radio}$ $-$ SFR$_{\rm OIR SED fit}$; the combined model with this additional component added is shown as a dashed line.  (Note that because this component is, by definition, completely obscured in the optical bands, it does not contribute significantly to our photometry outside the radio bandpass and its contribution can be calculated analytically without re-running the SED fit.)

Since luminous and star-forming galaxies are also often very dust-obscured, the afterglow properties of these events in relation to other GRBs (i.e., whether these bursts were optically dark or bright) are also of relevance.  We generally rely on previous analysis of the afterglows of these events from the literature, supplemented by our own reanalysis where necessary.

\subsection{GRB 051006}

GRB 051006 was a relatively weak \emph{Swift} burst with a faint X-ray afterglow.  Ground-based follow-up was limited and none of it was conducted less than six hours after the burst occurred.  Three optical sources near the burst location were initially identified as afterglow candidates by \cite{GCN4064}, none of which were reported to fade or vary \citep{GCN4068,GCN4089,GCN4094}.  This event may be a dark burst, although unfortunately without deeper, earlier, or redder-wavelength imaging it is impossible to know if this event was actually dust-extinguished to a significant extent.  While we classify it as a dark burst for the purpose of subsequent discussions, in this case the assignment is not definitive.

The refined XRT position \citep{Evans+2009} includes only source ``B'' of \cite{GCN4064}, which has therefore been identified as the probable host galaxy \citep{Hjorth+2012}, an association strengthened by its characterization as distant ($z=1.059$) and highly star-forming \citep{Jakobsson+2012}. The galaxy is easily detected in every filter in which we have observed it from the near-UV to 4.5 $\mu$m.  Modeling of the SED shows a moderately red optical (rest-frame UV) continuum indicative of a luminous, moderately extinguished galaxy.  We derive a best-fit extinction column of $A_V \sim 1.7$ mag and a star-formation rate of $\sim$100 $M_\odot$yr$^{-1}$.  (While a very low uncertainty on this value is inferred from our SED modeling, the exact SFR value depends on the assumed star-formation history; post-starburst models with lower instantaneous SFRs also give a good fit to the data but were disallowed.)  The mass of the older stellar population converged to zero in our fits, and the stellar population is dominated by the current episode with an intermediate age of about 100 Myr.

The UV/optical star-formation rate is comparable to, and possibly even in excess of, what we infer from the radio observations ($\sim$50 $M_\odot$yr$^{-1}$), indicating that this system is a luminous, star-forming galaxy but not an extreme SMG and it does not harbor large amounts of heavily-obscured star-formation.

\subsection{GRB 060814}

GRB 060814 is a prominent dark burst, with clear detections of an afterglow in $K$-band but not in any bluer filters \citepeg{GCN5455,Campana+2007}, unambiguously indicating an origin from an obscured region (but not an \emph{extremely} obscured region, else the $K$-band afterglow would also be undetectable: a reasonable bracket on the dust column is $1.8 < A_V < 8$ mag).

The host galaxy of GRB 060814 is discussed in some detail in \cite{Perley+2013a} as well as in \cite{Hjorth+2012} and \cite{Jakobsson+2012}.  It is well-detected in every optical/NIR band and is also one of the physically largest GRB hosts known, showing a complex, distorted morphology with an angular extent of $\sim 2.5$ arcsec (21 kpc at the spectroscopic redshift of $z=1.923$) that makes it easily resolved even in ground-based optical observations (Figure \ref{fig:images}; note that, as noted by \citealt{Jakobsson+2012}, the southwest source visible in the VLT image is a foreground object at $z=0.84$ but the remainder of the emission is from the host).  Our SED modeling indicates a very large star-formation rate of approximately 200 $M_\odot$yr$^{-1}$ after correction for dust extinction.  

The 3 GHz flux density of $\sim$15 $\mu$Jy measured for this source translates to a star-formation rate of 260 $M_\odot$yr$^{-1}$ at this redshift.  This is clearly within the range of ULIRGs and SMGs---although, again, the fact that this is in agreement with the optical SFR indicates that the galaxy is not actually optically thick but simply at the high-luminosity end of the population of ordinary, moderately-obscured galaxies.  However, this galaxy is clearly in a starburst phase, given its modest mass ($\sim$1.5$\times10^{10} M_\odot$) and highly distorted morphology.

A second source is also (marginally) detected in the VLA image a few arcseconds to the northwest of the host galaxy.  While there is no counterpart at this position in the $R$-band image shown in Figure \ref{fig:images}, a very red source is clearly detected there in the NIR imaging (HST WFC3-IR, $K$-band, and IRAC).  No redshift is available for this object, but given the relative rarity of very red objects and of faint radio sources it is possible that it may also be associated with the host system in some way.

\subsection{GRB 061121}

Unlike the other events in this list with radio detections, GRB 061121 was definitely not a dark burst.  Its afterglow was very bright in the optical and ultraviolet at early times, was detected in all UVOT filters including UVW2 (which partially overlaps the Lyman limit at the host redshift), and its optical spectral index was relatively normal \citep{Page+2007}, consistent with an unextinguished event.  These are all typical properties of the optically-bright GRB afterglows commonly studied (and targeted for host follow-up) in the pre-\emph{Swift} era.

Nevertheless, the event is hosted within a very luminous, radio-detected galaxy.  The 17$\mu$Jy source detected in the VLA image indicates a star-formation rate of 160 $M_\odot$yr$^{-1}$ (though with a substantial statistical uncertainty, since the RMS noise level of 5.5$\mu$Jy is higher than average.)   The host galaxy of this event is well-detected in all optical/NIR bands; the color is fairly blue and the inferred optical extinction is modest ($A_V \sim 0.4$ mag).  The UV/optical star-formation rate is 30 $M_\odot$yr $^{-1}$.

A secure interpretation of this system is somewhat complicated by the weak detection: nominally, the radio and optical star-formation rates are inconsistent by a factor of $\sim$5, but the radio measurement itself is uncertain by a factor of $\sim$2 simply due to the marginal detection and  additional systematic uncertainties underlie the star-formation rate derivations from UV and radio fluxes.  Nevertheless, taking our current measurement at face value indicates that a (modest) additional optically-thick starburst is present in the galaxy.  However, as the GRB itself shows no evidence of significant obscuration, GRB\ 061121 must have occurred in the relatively unobscured regions of the galaxy seen in our optical observations and \emph{not} from the optically-thick component.

A second source with similar flux level similar that of the host galaxy is present approximately 4.5$\arcsec$ to the northwest.  However, there is no optical counterpart at this location and the probability of chance occurrence of a noise peak at this flux level within a 4.5$\arcsec$ radius search region is significant, so we do not currently regard this object as real.

\subsection{GRB 070306}

GRB 070306 was a heavily obscured gamma-ray burst \citep{Jaunsen+2008}; its afterglow was detected only in $K$ and $H$ bands and its color between these filters was very red (consistent with $A_V \sim 5.5$ mag of extinction.)  The optical/IR host-galaxy observations of this target are all the same as previously presented in \citealt{Perley+2013a} (which in turn incorporated data from the earlier studies of \citealt{Jaunsen+2008} and \citealt{Kruehler+2011}) and despite somewhat different modeling assumptions we derive nearly identical parameters, including a very modest UV/optical star-formation rate of $\sim15$ $M_\odot$yr$^{-1}$.  The radio (and Herschel-) inferred dust-unbiased star-formation rate for this source is unambiguously much larger: approximately $\sim140$ $M_\odot$yr$^{-1}$, indicating the presence of substantial heavily-obscured star-formation.  The large column towards the GRB is consistent with an interpretation in which the burst itself actually occurred in the dominant, heavily obscured part of the host.

The best-fit stellar mass of this galaxy is the largest of the four radio-detected hosts in our sample ($\sim5 \times 10^{10} M_\odot$).   HST WFC3-IR imaging is available for this source \citep{Perley+2013a}; while the galaxy is resolved there are no obvious signs of an ongoing major merger at the resolution of the IR camera, though the galaxy does show some asymmetry (an extension to the northwest but not the southeast) and two much fainter companions (possibly, extended spiral arms) are present nearby.

\section{Notable Non-Detections}
\label{sec:nondetections}

Since the radio star-formation rates inferred for two of our detections are close to the values inferred from UV/optical observations of the same galaxies (ruling out significant optically-thick star formation), we also examined our sample to determine if any of our \emph{nondetections} imply radio luminosities close to our measured limits, which would also rule out additional optically-thick star formation.    The available TOUGH data only provide two filters ($R$ and $K_s$ bands), which are not adequate to evaluate the dust-corrected UV star-formation rates of these sources (dust corrections are critical even for reasonably optically thin galaxies).  However, in two cases additional data are available that suggest star-formation rates comparable to our radio limits, discussed below.

\subsection{GRB 060306}

The host galaxy of GRB 060306 was studied in \cite{Perley+2013a}.  Its optical/IR SED indicates a very luminous and dust-obscured galaxy; we previously derived an estimated star-formation rate of 245$^{+130}_{-67}$ $M_\odot$yr$^{-1}$ at the favored redshift of $z=1.55$ \citep{Perley+2013a}.  Our limiting radio star-formation rate of SFR $<$260 $M_\odot$yr$^{-1}$ is not inconsistent with this figure, but indicates that the galaxy---while probably a ULIRG, based on the bolometric luminosity implied by our fit---contains minimal optically-thick star formation.   This conclusion is similar to the one reached for this system from our previous study of dark GRB hosts with the VLA using C-band observations \citep{Perley+2013b} but significantly more constraining.  This target is at a declination near the satellite belt and the observation was strongly impacted by satellite RFI; had we achieved similar RMS sensitivities during our VLA observation of this target as were achieved for most of the other fields, we predict that the host would have been detected.

\subsection{GRB 060218}

GRB 060218 is by far the lowest-redshift event in the sample and represents something of a special case: it is extremely optically underluminous compared to almost any class of galaxy, is physically very small ($\sim$1 kpc), and clearly has little in common with the LIRG-like systems which our survey otherwise targets.  Nevertheless, given the close proximity of this galaxy (redshift $z=0.033$) the radio limit reaches very low star-formation rates.  This source was observed only in A-configuration and the beam (0.6\arcsec\ FWHM) is significantly smaller than the full optical extent of the host, but the core of the galaxy (in which the large majority of the star-formation is occurring) is very compact: \cite{Kelly+2014} estimate a half-light radius of only 0.31$\arcsec$ from HST imaging, so at least half of the galaxy's star-formation is contained within one VLA beam.   Conservatively allowing for a factor of $\sim$2 loss in sensitivity for this reason, our observations place a limit on the radio-derived SFR of $<$0.04 $M_\odot$yr$^{-1}$, similar to the optically-derived value of 0.03--0.04 $M_\odot$yr$^{-1}$ \citep{Wiersama+2007,Levesque+2010}.  Radio observations of very young, dwarf starburst galaxies often underpredict the optically derived star-formation rates because the ongoing starburst has not had sufficient time to produce enough supernova remnants to accelerate the electrons which produce the radio continuum \citep{Roussel+2003}, so a mild discrepancy would not be surprising for this system.

\subsection{Stacked Observations}

To provide a limit on the \emph{average} flux density of the host galaxies in our sample, we summed all 28 images in which no significant host emission was detected, using the host position from the VLT images as the alignment point.  Both a direct sum and a sum weighted using the inverse of the RMS were performed.  In neither case is significant emission detected at the position of the stacked host galaxies.  We measure an averaged flux density of 1.77 $\pm$ 0.85 $\mu$Jy in the unweighted stack, or 1.47 $\pm$ 0.77 $\mu$Jy in the weighted stack.  At the mean redshift of the stacked sample of $z=1.54$, this corresponds to a 2$\sigma$ limiting average star-formation rate of 41 $M_\odot$ yr$^{-1}$, although we caution that the flux-SFR conversion is a strong function of redshift and significant evolution is expected in the population across this period.

\begin{figure}
\centerline{
\includegraphics[scale=0.60,angle=0]{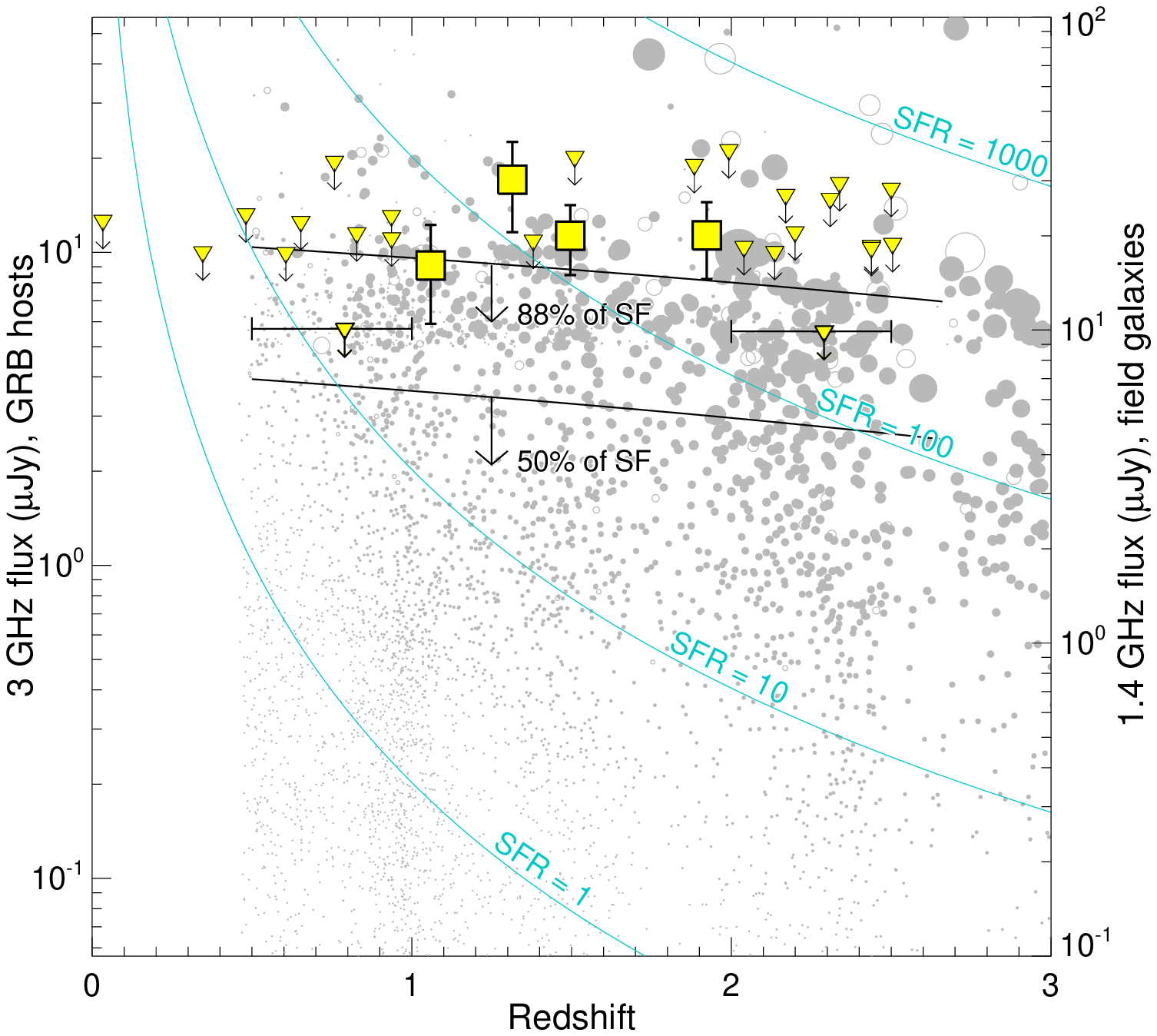}} 
\caption{3 GHz radio fluxes (or 3$\sigma$ limits) of TOUGH galaxies from the VLA (yellow data points), compared to measured or predicted radio fluxes from galaxies inside GOODS-N (grey circles).  To calculate the fluxes of the field galaxies, we use the reduced VLA map of this field presented by \cite{Morrison+2010} to directly match against the galaxy catalog of \cite{Kajisawa+2011}, scaling the 1.4 GHz flux density from that survey to 3 GHz using an average spectral index of $\alpha = 0.75$.  For sources which are not detected, we calculate the predicted radio flux density based on the star-formation rate estimates in \cite{Kajisawa+2011}.  Galaxies with hard X-ray detections are excluded (open circles).  We also evaluate the fraction of cosmic star-formation occurring in detected radio galaxies using this method by directly summing the star-formation rates of all galaxies above and below 10$\mu$Jy (see text for details).  The solid lines show a linear fit to the redshift-dependent 50th and 88th percentiles.  We also show the limits resulting from stacks of our nondetections at $0.5 < z < 1.0$ and $2.0 < z < 2.5$.}
\label{fig:radioflux}
\end{figure}

Somewhat more informative limits can be determined by stacking in limited redshift intervals.  Using the same procedure as in the previous paragraph, we stacked the images of 6 fields at $0.5 < z < 1.0$ and 9 fields at $2.0 < z < 2.5$; in the weighted stacks we do not detect a source in either case with nominal fluxes of $1.05\pm1.55$ $\mu$Jy and $1.60 \pm 1.33$ $\mu$Jy, respectively.  Equivalent 2$\sigma$ limiting star-formation rates are SFR $<$11.7 $M_\odot$yr$^{-1}$ at $z=0.79$ and $<$143 $M_\odot$yr$^{-1}$ at $z=2.29$.  These binned limits are shown in Figure \ref{fig:radioflux}.

\section{Population Analysis and Discussion}
\label{sec:demographics}

\subsection{Detection Rate and Connection to Cosmic Star-Formation}
\label{sec:detrate}

The successful detection of four objects with SFRs ranging up to $\sim$300 $M_\odot$ yr$^{-1}$ clearly indicates that GRBs can form in extremely luminous galaxies in at least some circumstances, while at the same time the low overall detection rate (approximately 15\%: 4 out of 24 low-RMS fields at known $z<2.5$, or 4 out of 29 including unknown-$z$) indicates that most GRB hosts are not systems of this type.  While this qualitative observation obviously rules out the most extreme models (e.g., in which GRBs are never produced by very luminous galaxies, or are produced only within very luminous galaxies), we wish to be more quantitative: is the observed detection rate above, below, or consistent with the detection rate \emph{expected} if GRBs neither prefer nor avoid the Universe's most luminous galaxies?  

\begin{figure}
\centerline{
\includegraphics[width=3.4in,angle=0]{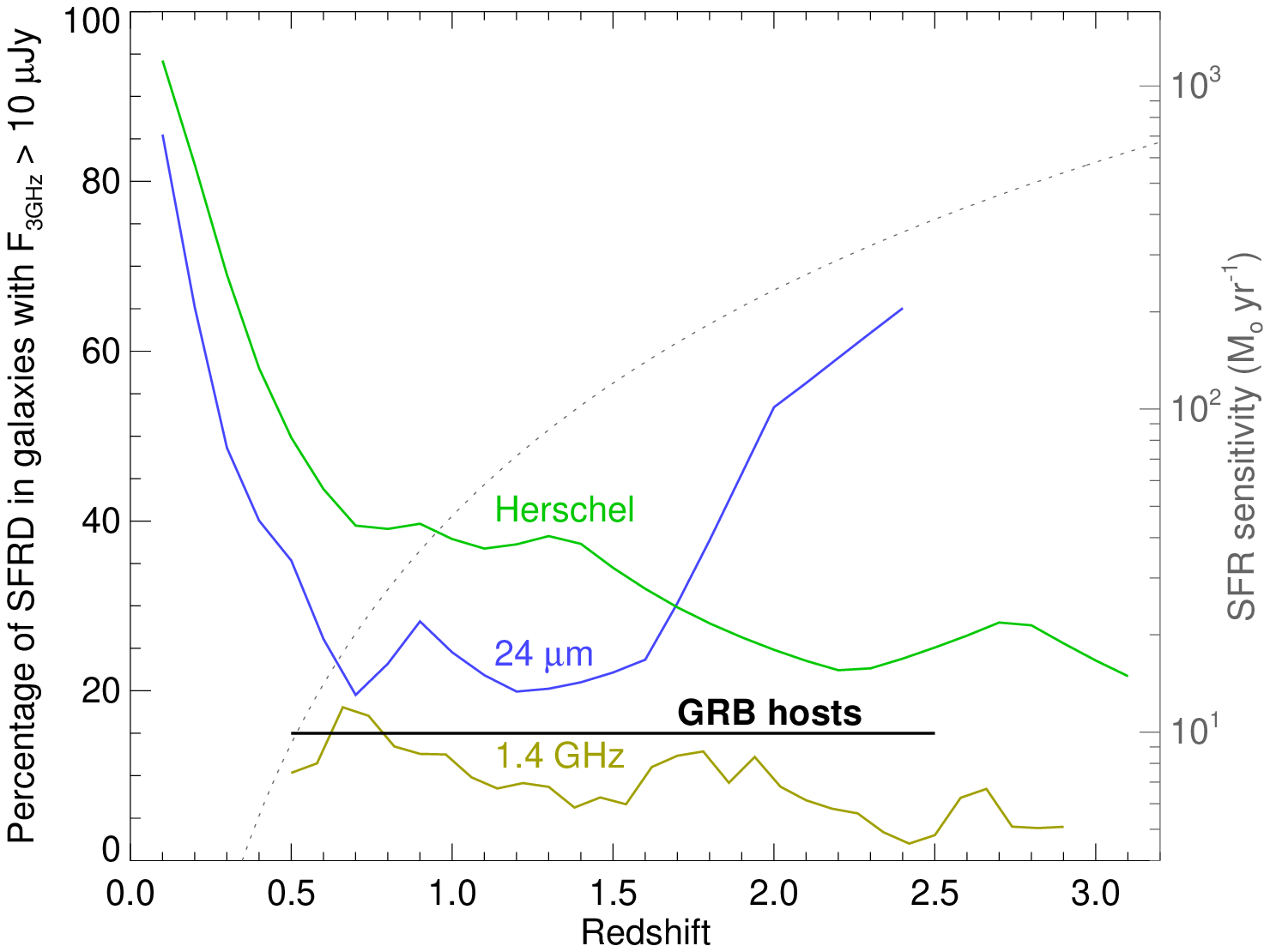}} 
\caption{Fraction of star-formation detectable to the approximate average sensitivity of our EVLA survey ($F_{\rm 3GHz} > 10 \mu$Jy) as a function of redshift, as calculated by three different means: integrating the Herschel-PACS luminosity functions of \cite{Gruppioni+2013}, integrating the Spitzer-MIPS luminosity functions of \cite{PerezGonzalez+2005}, and directly matching sources detected at 1.4 GHz in \cite{Morrison+2010} to the catalogs of \cite{Kajisawa+2011}.  Between approximately $z=0.5$ and $z=2.5$, radio observations probe about the same fraction of cosmic SFR with little variation with redshift, even though the \emph{absolute} star-formation rate sensitivity limit (right axis and dotted line) varies strongly with redshift.  Our GRB detection fraction is shown as a single bin, since our sample size is not yet large enough to provide redshift-resolved constraints.}
\label{fig:sfrpct}
\end{figure}

We followed two different approaches to predict the expected numbers under this SFR-tracing null hypothesis.   First, we numerically integrated long-wavelength luminosity functions from both Spitzer (the 24$\mu$m luminosity functions of \citealt{PerezGonzalez+2005}, converted to total infrared luminosity), and Herschel (the bolometric IR luminosity functions of \citealt{Gruppioni+2013}) to determine the quantity of star-formation occurring in galaxies with SFRs large enough to be detected in our survey at each redshift, using 10 $\mu$Jy as a characteristic flux limit of our survey.  Specifically, we calculate the quantity: $$f_{\rm det}(z) = \frac{ \int_{L_{\rm limit}(z)}^{\infty} L \times \phi_L(L,z) {\rm d} L }{ \int_{0}^{\infty} L \times \phi_L(L,z) {\rm d} L }$$  
Here, $L_{\rm limit}$ is the total IR luminosity of a star-forming galaxy with $F_{\rm 3 GHz}$ = 10 $\mu$Jy (our radio flux limit) and $\phi_L(L,z)$ is the IR luminosity function as taken from the literature.\footnote{Since it only examines dust-reradiated light, this procedure is insensitive to \emph{unobscured} star-formation and likely overestimates the fraction of star-formation in very luminous galaxies, but as most of the UV light in even ``normal'' galaxies is obscured at high-redshift \citepeg{Meurer+1999} the correction factor is likely to be small.  Based on the estimates of \cite{Oesch+2010}, the dust-\emph{uncorrected} UV star-formation rate density is not more than 20\% of the FIR-based star-formation rate density at any redshift relevant to this study.}
For the Herschel-based luminosity functions of \cite{Gruppioni+2013} the radio-detectable fraction drops gradually from 40\% to 25\% over our redshift range of interest ($0.5 < z < 2.5$).  The Spitzer-based fraction is flat at 20\% between $0.5 < z < 1.5$; it then \emph{rises} sharply above $z>1.5$ (Figure \ref{fig:sfrpct}) although 24\,$\mu$m-based luminosities are highly uncertain at these redshifts \citepeg{Papovich+2007,Elbaz+2010,Nordon+2010}.

Taking a more direct approach, we also calculated a sum of the actual inferred SFRs of all radio-detected galaxies from GOODS-N, using the redshift and SFR catalogs of \cite{Kajisawa+2011} matched with detected sources in the 1.4~GHz radio map of \cite{Morrison+2010}, which has similar resolution and sensitivity as our S-band observations.    We independently measured 1.4 GHz flux densities of sources in this map using the same procedure as for our GRB hosts, and scaled these values to 3 GHz using an average spectral index of $\alpha = 0.75$; galaxies with hard X-ray detections were excluded as AGN-dominated.  (The scaled, predicted fluxes from this method are shown as gray circles in \ref{fig:radioflux}.)  We then calculate the detectable fraction:
$$f_{\rm det}(z_{\rm min},z_{\rm max}) = \frac{ \sum_{{\rm galaxies\ with\ }F > 10 \mu{\rm Jy}} {\rm SFR}}{\sum_{\rm all\ galaxies} {\rm SFR}}$$
where both sums are restricted to galaxies over the range $z_{\rm min} < z < z_{\rm max}$.   SFR estimates are taken from the ``IR+UV'' column of \cite{Kajisawa+2011}, which uses the sum of UV and 24$\mu$m SFRs for MIPS-detected galaxies or the dust-corrected UV SFRs otherwise.   Since the GOODS-N field is small this method is more affected by small-number statistics than the Herschel/Spitzer maps, but the overall result is similar, between 5--15\% on average between $z=0.5-2.2$.

Note that despite the fact that there is rapid cosmic evolution with redshift over this range (the characteristic luminosity of galaxies increases by over an order of magnitude) and strong variation in sensitivity with redshift in the radio bands, these two effects largely cancel out, and our survey is sensitive to about the same \emph{proportion} of cosmic star-formation at every redshift:  specifically, at \emph{any} redshift between approximately $z=0.5$ and $z=2.5$, the galaxy population at or above our sensitivity limit is responsible for $\sim$5--30\% of the cosmic SFRD at that time.  
Therefore, regardless of the actual redshift distribution of the sample (as long as it is within this range), for an unbiased star formation tracer we expect a detection rate of 5--30\%.  

This range agrees quite well with our measured fraction of 15\%, so our radio data alone are consistent with the hypothesis that GRBs represent unbiased tracers of star-formation.   However, note that the prediction is systematically uncertain to a factor of $\sim$2--3 depending on the comparison study, and our own measurement is statistically uncertain to within a factor of $\sim$2--3 (\S \ref{sec:lumdist}), so this is not yet a highly constraining test on its own.  Fortunately, we know much more about these galaxies than simply their radio fluxes and star-formation rates and can apply additional, more constraining tests using additional parameters; we will return to this issue in \S \ref{sec:masses}.

We note that our detections cover a wide redshift range ($1.06 < z < 1.92$) and show no correlation between the observed flux and redshift, in support of our assertion that the detection rates and radio fluxes for the most luminous star-forming galaxies should be approximately constant with redshift.  We do not detect any galaxies at $z<1$ or $2 < z < 2.5$ within our sample, but given the small sample size this is not likely significant; indeed, previous targeted surveys have detected hosts at radio wavelengths at both higher and lower redshifts.

\subsection{Intrinsic Flux and Luminosity Distributions}
\label{sec:lumdist}

The convenient scaling between sensitivity and characteristic luminosity discussed above means that the detection fraction alone provides perhaps the most informative constraint on the degree to which GRBs prefer or avoid the most luminous galaxies, as long as it is understood that ``most luminous'' is treated as a \emph{relative} statement (versus other galaxies at the same redshift) as opposed to an absolute one (e.g., above some fixed luminosity threshold).\footnote{There are physical reasons why the relative definition may indeed be more relevant than an absolute one:  the star-formation rates of all types of star-forming galaxies appear to scale up with redshift in similar ways, such that luminous LIRGs and ULIRGs at $z\sim0$ are primarily mergers \citep{Sanders+1996,Veilleux+2002} whereas galaxies of similar luminosity at $z\sim1-2$ are primarily normal disks; major mergers instead produce extreme ULIRGs and HyLIRGs at these redshifts \citep{Chapman+2003b,Fadda+2010,Kartaltepe+2010b,Kartaltepe+2012}.}  Still, it is informative to produce formal constraints on the intrinsic flux distribution inferred from our survey and on the luminosity distribution as a function of redshift.

While we have treated our detection fraction as a single value above a single flux point, in reality our sensitivity limits do vary from field to field (from 7.1 $\mu$Jy for 070306 to 18.6 $\mu$Jy for 060306 and 070621, at 2.5$\sigma$).   Translating our results into a statistical constraint on the actual fraction with flux density above a specific value therefore requires a prior on the shape of the flux density distribution of the underlying population.  We assume it follows a single flux-weighted Schechter function (${\rm d}N/{\rm d}S \propto S \times (S/S*)^\alpha {\rm e}^{-S/S*}$, with $\alpha=-1.1$ and $S*$ a free parameter)---a reasonable assumption if GRBs approximately trace star-formation given the redshift-independence of the characteristic flux as described in the previous paragraph---although the results are not particularly sensitive to the choice of model.  Based on a simple Monte-Carlo analysis using the distribution of 2.5$\sigma$ sensitivity limits achieved by our survey and including the effects of measurement noise, we calculate that, intrinsically, between 9 and 23 percent of all $z<2.5$ GRB hosts have $S_{\rm 3 GHz} > 10 \mu$Jy at 90\% confidence.

We do not have enough detections to constrain the flux density function shape in a meaningful way, but it is worth noting that we detect \emph{no} host galaxies with a flux density greater than 20$\mu$Jy at any redshift, so the luminosity function is at least moderately steep (as indeed is expected for the high-luminosity exponential tail of a Schechter-like population).   This observation produces some tension with the pre-\emph{Swift} study of \cite{Berger+2003}, which reported a detection rate of 20\% at a 2.5$\sigma$ radio sensitivity level of 20--30 $\mu$Jy in X-band (equivalent to 40--60 $\mu$Jy in S-band for $\alpha=0.75$).  The distribution of redshifts for the targets in their survey, and for their reported detections, is similar to what is seen in our own.  We suspect that some of their reported detections may have either been noise fluctuations or affected by late-time contribution from the GRB afterglow, and/or that their detection of several intrinsically rare, luminous objects in such a small sample was a statistical fluke.

Again assuming that the ratio of host luminosity sensitivity and characteristic star-forming galaxy luminosity can be assumed to be constant over our redshift range, the flux-distribution constraint can be translated to an equivalent SFR-distribution constraint at any redshift within a factor of $\sim$2 simply by calculating the equivalent limiting star-formation rate for our 10$\mu$Jy limiting flux at each redshift.  For example, we estimate that 9--23\% of GRB hosts at $z\sim0.6$ have ${\rm SFR}>14 M_\odot$, 9--23\% of hosts at $z\sim1.0$ have ${\rm SFR}>50 M_\odot$, 9--23\% of hosts at $z\sim1.5$ have ${\rm SFR}>125 M_\odot$, and 9--23\% of hosts at $z\sim2.0$ have ${\rm SFR}>250 M_\odot$.

\subsection{The Stellar Masses of GRB-Selected Submillimeter Galaxies}
\label{sec:masses} 

In the study of \cite{Perley+2013a} we examined the influence of several parameters on the rate at which a galaxy produces GRBs relative to stars in general (the GRB production ``efficiency'', $\epsilon \equiv R_{\rm GRB}/{\rm SFR}$).   The total star-formation rate of a galaxy had no discernable effect on its efficiency for GRB production over the redshift range we examined ($z \sim 0.5-3$).  The consistency of the number of radio detections in our TOUGH-VLA sample with the expected number for an unbiased tracer (\S \ref{sec:lumdist}) can be seen as an extension of this conclusion up to the highest star-formation rates.  On the other hand, we previously found that mass had a strong effect (GRBs in the most massive galaxies are much rarer than expected given the contributions of such galaxies to overall cosmic star-formation), so it is relevant to investigate the masses of the luminous galaxies detected in this effort as well.

\begin{figure}
\centerline{
\includegraphics[width=3.4in,angle=0]{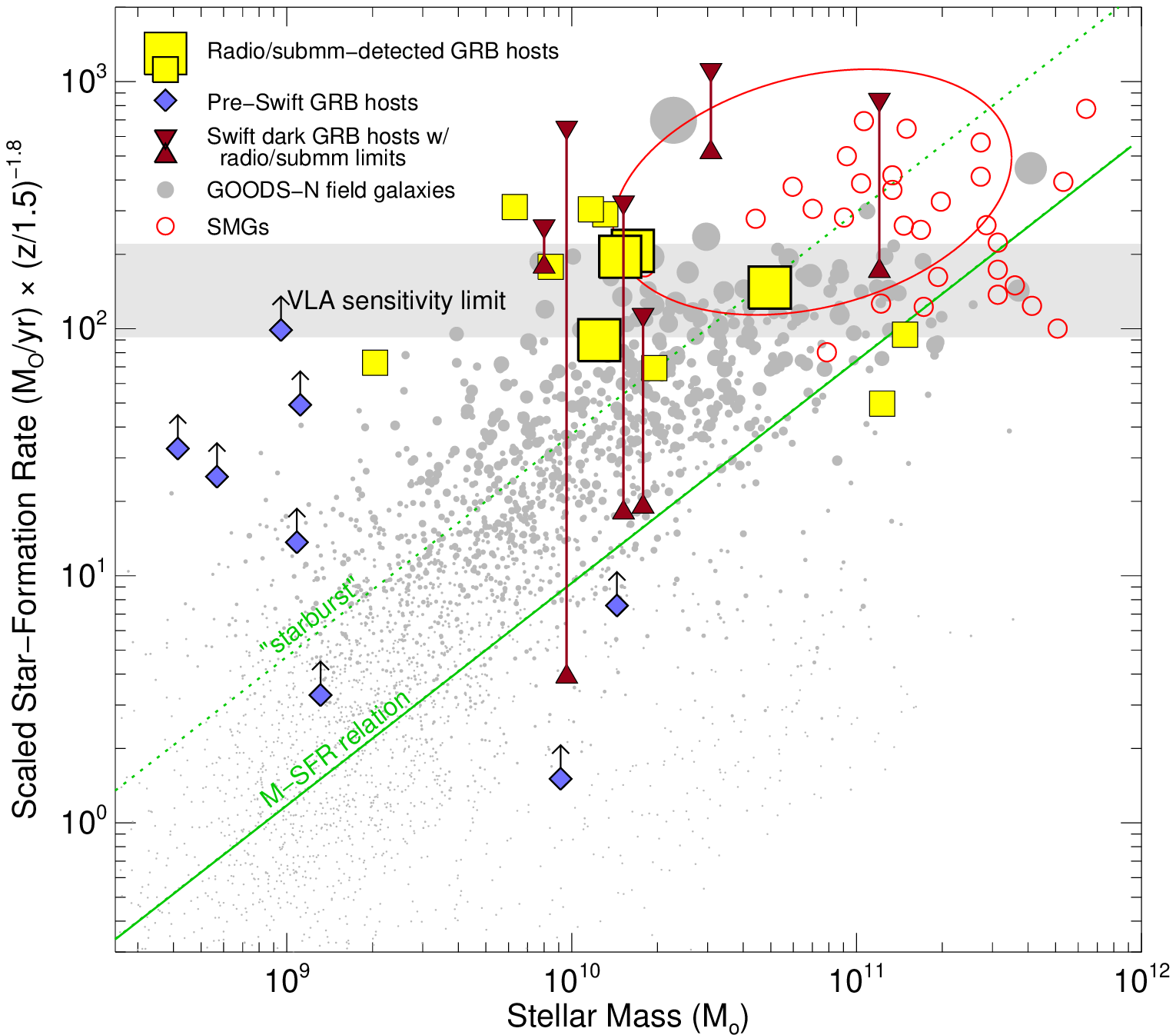}} 
\caption{Stellar masses and star-formation rates of GRB host galaxies and field galaxies from our sample and from the literature at $0.8 < z < 2.2$.  In addition to the four radio-detected targets presented here (large squares), we plot GRB hosts with FIR/submillimeter/radio detections from \cite{Michalowski+2008} and \cite{Hunt+2014} (small squares), as well as events from \cite{Perley+2013b} which were \emph{not} detected at radio wavelengths, showing the range between the UV/optical SFR as a minimum and the radio SFR as a maximum in the form of vertical lines.  Pre-\emph{Swift} GRB hosts are also shown, plotted as lower limits since sensitive radio/submillimeter limits are generally not available.  For the GOODS-N field galaxies \citep{Kajisawa+2011} we take 24$\mu$m SFRs when available or dust-corrected UV SFRs otherwise, correcting by an IMF adjustment factor of 1.6 to convert from Salpeter to Chabrier/Kroupa.  To minimize the impact of redshift evolution and varying radio luminosity-sensitivity over this broad interval we have scaled all star-formation rates by a factor of $(z/1.5)^{-1.8}$.  The galaxy $M-SFR$ relation at $z=1-2$ \citep{Elbaz+2007,Daddi+2007} is also shown; galaxies with sSFRs a factor of five above this relation are defined as ``starbursts'' \citep{Rodighiero+2011}.  Compared to field galaxies of comparable luminosities and compared to SMGs (small circles indicate individual systems from \citealt{Michalowski+2010}; the large red ellipse represents the region of parameter space inhabited by SMGs in \citealt{Gruppioni+2013}), the most luminous GRB hosts have unusually low stellar masses.}
\label{fig:masssfr}
\end{figure}

Our optical/NIR SED modeling provides firm estimates of the stellar masses of all four-radio detected galaxies (\S \ref{sec:detections})\footnote{Of course, stellar masses in this and all other studies are subject to systematic uncertainties related to the input assumptions.  As discussed earlier, our modeling assumes a Chabrier IMF and a two-population star-formation history, which are thought to be the most realistic assumptions for fitting galaxies of this type \citep{Michalowski+2012a,Michalowski+2014b}.  However, even when employing alternative star-formation histories we derived consistent results, and in this analysis we scale all comparison studies to the same IMF.}, enabling us to evaluate whether or not their mass distribution is consistent with that which would be expected for the most luminous members of a star-formation-sampled galaxy population.   Results are plotted in Figure \ref{fig:masssfr}; note that due to the strong cosmic evolution in the mass-SFR relation over the redshift interval spanned by our sample we have applied a redshift scaling factor of $(z/1.5)^{-1.8}$ to all star-formation rates\footnote{The power-law index of $-1.8$ is derived by comparing the mass-SFR relation at $z\sim1$ from \cite{Elbaz+2007} to that at $z\sim2$ from \cite{Daddi+2007}.  After this adjustment these two relations lie along the same solid green line, shown on the plot.} to remove this effect from our analysis.  The masses of our radio-detected hosts (large yellow squares) are well below the masses of almost all high-luminosity (scaled SFR $>100$ $M_\odot$yr$^{-1}$) galaxies in the GOODS-N field as estimated by \citealt{Kajisawa+2011}, with the exception of GRB 070306 whose stellar mass is fairly typical for a luminous galaxy.

Because the GOODS-N field is small, galaxies with the highest masses and star-formation rates are not well-sampled and this comparison may be biased by cosmic variance.  However, wide-area FIR and submillimeter surveys select galaxies out of the underlying field population in a manner similar to the way in which our four detections were selected out of the parent TOUGH sample, and provide a comparison sample over a much larger volume suitable to evaluate the GRB-SFR connection within luminous galaxies in more detail.  In Figure \ref{fig:masssfr} we plot both a sample of submillimeter-selected SMGs from \cite{Michalowski+2010} as well as a broad ellipse that encloses nearly every SMG in the Herschel studies of \citealt{Rodighiero+2011} (their Figure 1) and \citealt{Gruppioni+2013} (their Figure 15).

The stellar masses of submillimeter galaxies are subject to some systematic uncertainty, with the values derived varying by up to a factor of almost ten depending on the assumptions considered in modeling the stellar populations (in particular the star-formation rate history, choice of stellar templates, and IMF; see \citealt{Michalowski+2012a,Michalowski+2014b} for an in-depth discussion of this issue).  The offset of the \cite{Michalowski+2012a} SMGs and the Herschel samples in Figure \ref{fig:masssfr} is largely a reflection of this.   In our own modeling we have assumed a two-component stellar population, which tends to lead to \emph{higher} masses when applied to SMGs.  \footnote{We verified this explicitly by refitting all of the SMGs with UV-through-NIR host detections from \cite{Michalowski+2008} using the same procedure employed for our GRB hosts and derived masses consistent with the higher values that work.}  
However, even when compared to SMG mass catalogs derived using assumptions leading to lower masses, very few such galaxies are of masses comparable to the GRB hosts in our sample (three out of four of the GRB hosts in our sample have a mass below $2\times10^{10} M_\odot$, and SMGs below this value are extremely uncommon.)
\footnote{An additional systematic issue concerns the completeness and purity of the submillimeter and Herschel samples discussed here.  While recent Herschel papers claim very high redshift completeness (80-100\%; \citealt{Gruppioni+2013}), it is not inconceivable that some Herschel sources are incorrectly identified with bright counterparts in the Herschel beam when in fact they should be associted with fainter, lower-mass sources, leading to a bias in favor of larger masses and lower redshifts.  ALMA follow-up will be necessary to clarify this issue unambiguously, but since our survey targets a relatively nearby redshift range ($z=1-2.5$) at which counterpart matching is fairly secure we will treat the published Herschel/JCMT results at face value.}

Our sample size is small (4 hosts), but can be expanded by considering additional submillimeter/radio-detected GRB hosts from previous work (although these were not selected out of an unbiased GRB sample, and one of which was detected in radio but undetected at submillimeter wavelengths and could conceivably be afterglow-contaminated; \S \ref{sec:detrate}).  \cite{Michalowski+2008} analyzed the SEDs of four submillimeter- or radio-detected GRB hosts (all at $z=1-2$), and derived stellar masses less than $2 \times 10^{10} M_\odot$ (after IMF adjustment) for all four cases.  Furthermore, a few pre-Swift host galaxies with even lower stellar masses ($\sim10^9 M_\odot$) have measured UV dust-corrected star-formation rates indicating predicted radio fluxes similar to our detection level (Figure \ref{fig:masssfr}).  This result is fully consistent with the distribution observed within our own sample and provides additional support to the notion that, even within submillimeter galaxies and at $z\sim1.5$, GRBs exhibit a strong aversion against the most massive galaxies.  

GRBs in more massive submillimeter-like galaxies do exist, particularly among dark GRB hosts: combining the dark GRB samples of \cite{Perley+2013b} and \cite{Hunt+2014}, four galaxies with masses of $>5\times10^{10} M_\odot$ have been detected by Herschel or VLA.   However, it is important to keep in mind that these are not unbiased samples, but are selected out of the much larger parent pool of the hundreds of \emph{Swift} and pre-\emph{Swift} GRBs and subject to very specific selection criteria, explicitly favoring heavily reddened afterglows and IRAC-bright (massive) galaxies.  

Further long-wavelength observations of unbiased samples will be necessary to confirm this interpretation, but these results offer support to the notion that GRBs preferentially explode in low-mass galaxies (albeit without entirely excluding massive galaxies) across the entirety of the host luminosity range.

\subsection{GRBs from Optically Thick Super-Starbursts?}

Traditional submillimeter galaxies are extremely heterogeneous systems, with dense, intensely star-forming cores that dominate the bolometric output (and overall star-formation rate) as well as less-obscured outer components that dominate the UV/optical luminosity despite their minor contributions to the total energy budget.  Since the degree of obscuration to a GRB can be determined from afterglow observations, when a gamma-ray burst is localized to a submillimer galaxy it is often possible to crudely discern where within the host the GRB actually occurred: a GRB that is extremely obscured ($A_V > 5$ mag) likely occurred within the central, optically-thick starburst whereas a GRB showing little or no obscuration must have occurred in the optically-thin regions.   

The sites of GRBs within heterogeneous submillimeter galaxies represent an important indicator of the connection between GRB production and star-formation, since the dusty inner regions of these galaxies are likely to be much more metal-rich than the unobscured regions but are also much more intensely star-forming.   Different physical hypotheses for the origin of the variations in the GRB rate have opposing predictions for the rate of GRBs in dusty central starbursts:  if the GRB efficiency is governed primarily by metallicity we would expect GRBs to avoid the obscured regions; if the GRB efficiency is governed primarily by star-formation intensity then we would expect GRBs to be abundant in obscured regions.

Only two of the galaxies in our sample show evidence of strong dust heterogeneity: the hosts of GRB 061121 and GRB 070306.  These two sources offer a split verdict on the location of the burst within these hosts.  The bright, blue afterglow of GRB 061121 is clearly not heavily obscured: \cite{Page+2007} estimate between $A_V = 0.4-1.6$ mag depending on the assumed dust law and intrinsic spectral index.  This is consistent with the mean obscuration inferred from the host UV/optical observations in our SED modeling and clearly situates the event outside the heavily obscured star-forming region that our data (tentatively) indicate is also present.  On the other hand, GRB 070306 was extremely obscured and this event is quite consistent with having occurred within the dusty starburst region of a submillimeter galaxy.

Two events are not sufficient to give a definitive answer as to whether or not GRBs preferentially prefer or avoid any particular area of this type of galaxy.  Again, however, comparison to other literature results can offer a hint.  Among the four pre-Swift submillimeter host galaxies discussed in \cite{Michalowski+2008}---all of which showed evidence for dust heterogeneity as discussed above---three were optically bright and therefore unobscured; only one, GRB 000210, was consistent with an obscured origin.  While it is conceivable that this is a selection effect (few optically-obscured dark bursts were sufficiently well-localized in the pre-Swift era to perform a host galaxy search), dedicated studies of dark GRBs have failed to turn up many counterexamples:  all of the submillimeter or radio-detected galaxies of dark GRBs in the studies of \cite{Hunt+2014} and \cite{Perley+2013b} have corresponded to optically reddened galaxies in which the gap between the extinction-corrected UV and the FIR/radio star-formation rates is small.  This may be a selection effect on its own (in \citealt{Perley+2013b} we explicitly favored events with redder optical colors and higher inferred dust-corrected star-formation rates; \citealt{Hunt+2014} observed only hosts previously targeted with, and detected by, \emph{Spitzer}), but even so, additional analogs of GRBs 070306 and 000210 have proven remarkably elusive considering that the bolometric luminosity of obscured regions of typical submillimeter galaxies can exceed the luminosities of the unobscured regions by huge factors.

Larger, preferentially unbiased, campaigns to observe GRB host galaxies at submillimeter/radio wavelengths will be required to resolve this question unambiguously.  However, the small numbers of GRBs found to date within heavily obscured regions of very luminous, blue galaxies (two) in relation to the much larger number found in \emph{un}obscured regions of similar galaxies (as many as five, although c.f.\ our discussion in \S \ref{sec:detrate}) may point toward nonuniform efficiency for GRB production across the different parts of these hosts: preference for outer, unobscured regions and for blue, moderately-extinguished ULIRGs, and an aversion against the central, dusty starburst.

\section{Conclusions}
\label{sec:conclusions}

We have presented sensitive radio observations to search for host-galaxy synchrotron emission at the locations of 31 GRBs selected from the TOUGH host-galaxy survey.   Among 25 sources at known redshift, we securely detect three (and probably detect a fourth), indicating that approximately 15\% of all GRBs occur in the most luminous galaxies (those detectable at radio wavelengths).

As about 15\% of all cosmic star-formation occurs in galaxies of similar radio luminosity over this redshift range, our results suggest that GRBs are not particularly biased tracers of cosmic star-formation \emph{with regard to bulk galaxy star-formation rate}, neither markedly preferring nor avoiding galaxies with the highest luminosities.  However, the stellar masses of the systems we detect with the VLA are about an order of magnitude lower than those of field-selected star-forming galaxies of similar luminosities and redshifts.  Both of these results are consistent with the trends seen in \cite{Perley+2013a}, in which the GRB host population had significantly lower mass than an SFR-weighted field sample but matched the expected distribution of UV-based star-formation rates.

Why would GRBs appear to preferentially avoid massive galaxies, yet not avoid the most actively star-forming galaxies, when it is well-established that the most massive galaxies are, on average, the most-star-forming?   Our results suggest that the GRB efficiency, while depressed in massive galaxies, must also be \emph{enhanced} in galaxies with high \emph{specific} star-formation rates, allowing moderate-mass starburst galaxies to compensate for the underabundance of GRBs in non-starbursting high-mass galaxies.   This effect probably operates at the low-mass end as well, given the remarkable abundance of GRBs within low-mass, modest total-SFR, extremely high-sSFR galaxies (blue points in Figure \ref{fig:masssfr}, or Figure 13 from \citealt{Perley+2013a}; see also \citealt{Christensen+2004}, \citealt{CastroCeron+2006,CastroCeron+2010}, and \citealt{Svensson+2010}) which contribute almost negligibly to overall cosmic star-formation except at the highest redshifts.

The traditional hypothesis for the fundamental factor controlling the GRB rate is a metallicity effect, since such an effect is predicted theoretically in many models \citepeg{MacFadyen+1999, Yoon+2005} and several observational studies have found a strong trend for the GRB population to favor low-metallicity galaxies \citep{Modjaz+2008,Levesque+2010,Graham+2013}.  Since mass and metallicity are strongly correlated \citepeg{Tremonti+2004}, this naturally explains why GRB hosts also seem to favor low-mass galaxies.   Evidence for an additional dependence on specific star-formation rate (independent of that on mass) does not necessarily nullify this conclusion, since high-sSFR galaxies may be more metal-poor than low-sSFR galaxies of the same mass \citep{Mannucci+2010,LaraLopez+2010}---but the sSFR-metallicity trend is not particularly strong, so it remains to be seen whether the distribution of GRB host properties is truly consistent with a model in which the GRB efficiency depends only on metallicity and no other factors.   Direct measurements of the metalliticies of the systems studied here (and of other low-metallicity galaxies) should provide some insight:  are these galaxies actually significantly less chemically enriched than others of the same mass, or simply more active?  In the latter case, this would provide evidence for alternative scenarios in which another parameter instead of or in addition to metallicity governs the GRB efficiency, such as a variable IMF or binary interactions in dense clusters \citep{vandenHeuvel+2013}.  Some independent evidence for such a scenario is provided by the spatially resolved characteristics of GRB host galaxies: HST studies have shown that GRB hosts are more compact than core-collapse supernova hosts and that the GRB itself preferentially occurs in or near the brightest, densest regions of each galaxy \citep{Fruchter+2006,Svensson+2010,Kelly+2014,Michalowski+2014}.

While only four radio-luminous systems have been identified in an unbiased manner so far, the path to identifying more is now well-established: it is now possible to efficiently survey large numbers of fields with the VLA, and an extension of this technique could easily identify many new submillimeter/radio-luminous host galaxies in the next few years.   Our survey demonstrates the depths necessary to detect an appreciable fraction of GRB hosts out of an unbiased sample (10 $\mu$Jy at 3 GHz to identify $\sim$15\%, independent of the redshift of the host population being targeted as long as it is $z\lesssim2.5$) a limit achievable in only $\sim$2 hours of observation.  ALMA observations will be even more sensitive, reaching an order of magnitude fainter and probing GRB hosts beyond $z>3$, as has already been demonstrated by \cite{Wang+2012} for two objects.  A complete understanding of the entire GRB host population, from the least to the most luminous galaxies and from the nearby universe out to high redshifts, is clearly within reach.

\vskip 0.02cm

\acknowledgments

Support for this work was provided by NASA through Hubble Fellowship grant HST-HF-51296.01-A awarded by the Space Telescope Science Institute, which is operated by the Association of Universities for Research in Astronomy, Inc., for NASA, under contract NAS 5-26555.
The National Radio Astronomy Observatory is a facility of the National Science Foundation operated under cooperative agreement by Associated Universities, Inc.
The W. M. Keck Observatory is operated as a scientific partnership among the California Institute of Technology, the University of California, and NASA; the Observatory was made possible by the generous financial support of the W. M. Keck Foundation.  We extend special thanks to those of Hawaiian ancestry on whose sacred mountain we are privileged to be guests.  
This work is based in part on observations made with the  {\it Spitzer Space Telescope}, which is operated by the Jet  Propulsion Laboratory, California Institute of Technology, under a contract with NASA.  Partial support for this work was provided by NASA through an award issued by JPL/Caltech.  It is also based in part on observations with the NASA/ESA {\it Hubble Space Telescope}, obtained from the Space Telescope Science Institute.  STScI is operated by the Association of Universities for Research in Astronomy, Inc. under NASA contract NAS 5-26555.  These observations are associated with program GO-12949.
We thank F.~Owen for discussions regarding deep radio field surveys, C.~Casey for advice regarding the star-formation rates and masses of submillimeter- and Herschel-selected galaxy populations, and D.~A.~Kann for helpful comments.  We thank U. Rau for assistance with analysis of the GRB 061110A data set, and H.~Knutson, M.~Zhao, and J.~Curtis for acquiring the $J$ and $K_s$ imaging of GRB 051006.   We also thank the anonymous referee for helpful suggestions that improved the quality of the paper.  This manuscript was completed during the ``Fast and Furious: Understanding Exotic Astrophysical Transients'' workshop at the Aspen Center for Physics, which is supported in part by the NSF under grant No.\ PHYS-1066293.


{\it Facilities:} \facility{VLA, Keck-I:LRIS, Palomar:WIRC, Spitzer:IRAC, HST:WFC3}



\end{document}